\documentclass[aps,prb,floatfix,twocolumn,superscriptaddress,hyperref=pdftex]{revtex4-2}

\usepackage{amssymb}
\usepackage{hyperref}
\hypersetup{
    colorlinks,
    linkcolor={blue!80!black},
    citecolor={blue!80!black},
    urlcolor={blue!80!black}
}
\usepackage{graphicx}
\usepackage{placeins}
\usepackage{booktabs}

\usepackage{upgreek}
\usepackage{siunitx}
\usepackage{physics}
\usepackage{soul}
\usepackage{xcolor}
\usepackage{tabularx,multirow,array}
\usepackage{siunitx}
\usepackage{chemmacros}
\usepackage[normalem]{ulem}

\definecolor{SDEcolor}{rgb}{0.8, 0, 0.}
\newcommand{\SDEcomment}[1]{}
\newcommand{\Elsa}[1]{}

\begin{document}
\title{Semiconductor-ferromagnet-superconductor planar heterostructures for 1D topological superconductivity}
\author{Samuel D. Escribano}
\email[Corresponding author: ]{samuel.diaz@uam.es}
\affiliation{Departamento de Física Teórica de la Materia Condensada C5, Condensed Matter Physics Center (IFIMAC) and Instituto Nicolás Cabrera, Universidad Autónoma de Madrid, E-28049 Madrid, Spain}
\author{Andrea Maiani}
\affiliation{Center for Quantum Devices, Niels Bohr Institute, University of Copenhagen, 2100 Copenhagen, Denmark}
\author{Martin Leijnse}
\affiliation{Center for Quantum Devices, Niels Bohr Institute, University of Copenhagen, 2100 Copenhagen, Denmark}
\affiliation{Division of Solid State Physics and NanoLund, Lund University, 22100 Lund, Sweden}
\author{Karsten Flensberg}
\affiliation{Center for Quantum Devices, Niels Bohr Institute, University of Copenhagen, 2100 Copenhagen, Denmark}
\author{Yuval Oreg}
\affiliation{Department of Condensed Matter Physics, Weizmann Institute of Science, Rehovot 7610, Israel}
\author{Alfredo Levy Yeyati}
\affiliation{Departamento de Física Teórica de la Materia Condensada C5, Condensed Matter Physics Center (IFIMAC) and Instituto Nicolás Cabrera, Universidad Autónoma de Madrid, E-28049 Madrid, Spain}
\author{Elsa Prada}
\affiliation{Instituto de Ciencia de Materiales de Madrid (ICMM), Consejo Superior de Investigaciones Científicas (CSIC), E-28049 Madrid, Spain}
\author{Rub\'en Seoane Souto}
\affiliation{Center for Quantum Devices, Niels Bohr Institute, University of Copenhagen, 2100 Copenhagen, Denmark}
\affiliation{Division of Solid State Physics and NanoLund, Lund University, 22100 Lund, Sweden}

\date{\today}
\begin{abstract}
Hybrid structures of semiconducting (SM) nanowires, epitaxially grown superconductors (SC), and ferromagnetic-insulator (FI) layers have been explored experimentally and theoretically as alternative platforms for topological superconductivity at zero magnetic field. Here, we analyze a tripartite SM/FI/SC heterostructure but realized in a planar stacking geometry, where the thin FI layer acts as a spin-polarized barrier between the SM and the SC. We optimize the system's geometrical parameters using microscopic simulations, finding the range of FI thicknesses for which the hybrid system can be tuned into the topological regime. Within this range, and thanks to the vertical confinement provided by the stacking geometry, trivial and topological phases alternate regularly as the external gate is varied, displaying a hard topological gap that can reach half of the SC one. This is a significant improvement compared to setups using hexagonal nanowires, which show erratic topological regions with typically smaller and softer gaps. Our proposal provides a magnetic field-free planar design for quasi-one-dimensional topological superconductivity with attractive properties for experimental control and scalability.

\end{abstract}

\maketitle

\section{Introduction}
The interplay between superconductivity and magnetism in low-dimensional systems allows to engineer quantum phases absent in nature otherwise. Topological superconductors are paradigmatic examples, hosting Majorana-like quasiparticles at their boundaries or near defects. The exotic properties of these bound states, including their non-locality and non-abelian exchange statistics, have attracted a growing interest in the field~\cite{Alicea_RPP2012,Leijnse_Review2012,Aguado_Nuovo2017,Prada_review,Lutchyn_NatRev2018,Flensberg_NatMat2021}. In particular, they are ideal platforms for encoding and processing quantum information in a protected way~\cite{NayakReview}.

\begin{figure}[h!]
    \centering
    \includegraphics[width=1\columnwidth]{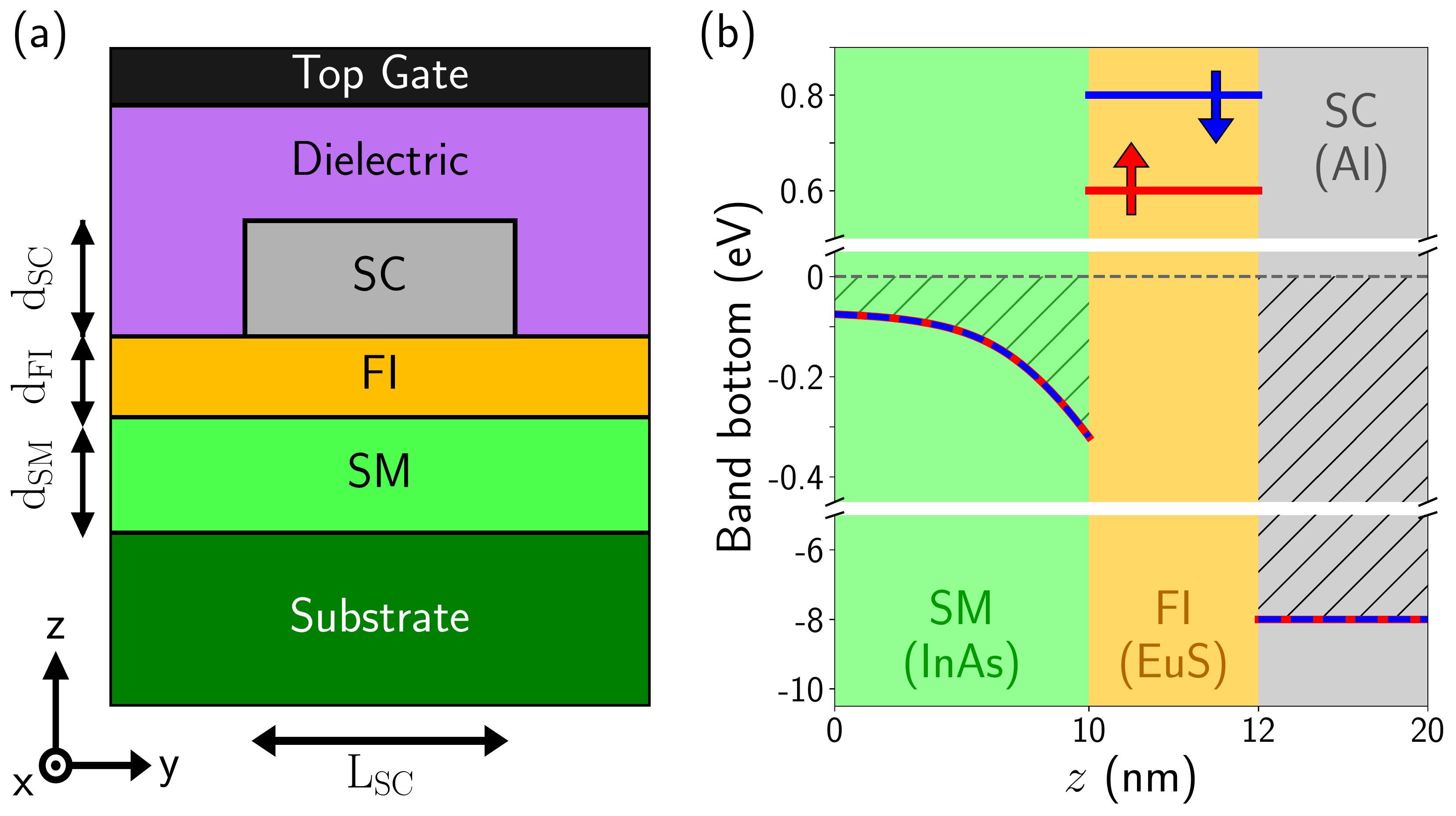}
    \caption{\textbf{Hybrid planar heterostructure.} (a) Sketch of the device studied in this work: 2D semiconductor (SM)/ferromagnetic insulator (FI)/superconductor (SC) heterostructure stacked in the $z$-direction and infinite in the $x$-direction. The substrate is insulating and typically composed of several stacked semiconducting layers. The top gate can be used to confine the wavefunction below the grounded SC. The thickness of the ferromagnetic insulator layer $d_{\rm{FI}}$ is varied to optimize the topological properties. (b) Schematics of the conduction-band bottom along the heterostructure stacking direction for a specific choice of materials (\ch{InAs}/\ch{EuS}/\ch{Al}) and representative geometrical parameters ($d_{\rm SM}=10$ nm, $d_{\rm FI}=2$ nm, $d_{\rm SC}=8$ nm and $L_{\rm SC}=100$ nm). Red and blue colors represent different spin directions, and the gray dashed line depicts the Fermi level.
    }
    \label{Fig1}
\end{figure}

Theory proposals suggested the onset of topological superconductivity in semiconductor (SM) nanowires with strong spin-orbit coupling when proximitized by a superconductor (SC)~\cite{Oreg_PRL2010,Lutchyn_PRL2010}. As an external magnetic field increases, the system undergoes a topological quantum phase transition, characterized by the closing and reopening of the superconducting gap. In the topological regime, sufficiently long wires feature zero-energy Majorana bound states at the ends.
Robust zero-bias conductance peaks compatible in principle with Majorana states have been measured in nanowires over the last decade~\cite{Mourik_science2012, Das_NatPhys2012, Deng_Science2016, GulNanoLett2017,Grivnin_NatCom2019}. Later works have shown zero-energy states also in two-dimensional (2D) SM/SC hybrids~\cite{Suominen:PRL17, Nichele_PRL2017,Hell_PRB2017, Fornieri_Nat2019}, an ideal platform for multi-wire designs with a measured high mobility~\cite{Kjaergaard_NatCom16, Lee_NanoLet19, Ahn_PRM2021}. However, the strong external magnetic field needed for the topological transition is detrimental to superconductivity and sets strict constraints on the device geometry, since the applied field needs to be oriented parallel to each wire. This is an obstacle for experiments showing Majorana non-abelian properties \cite{BeenakkerReview_20} and, ultimately, for topological quantum devices. Devices based on magnetic flux through full-shell nanowires \cite{Vaitiekenas_Science2020, Penaranda_PRR2020, Valentini_Science2021} and the phase difference in superconducting junctions \cite{Pientka_PRX2017, Lesser_JoPD2022, Banerjee_arXiv2022} are alternatives considered recently. However, these designs offer drawbacks for device scaling, due to their magnetic field direction sensitivity or the difficulty of controlling the phase difference between many superconductors.

In this context, ferromagnetic insulators (FIs) offer a way to solve the above problems by inducing a local exchange field on the SM nanowire by proximity effect, eliminating the need for external magnetic fields. Recent experiments in hexagonal nanowires partially covered by overlapping SC and FI shells showed the appearance of zero-bias conductance peaks  \cite{Vaitiekenas_NatPhys2020}, spin-polarized subgap states \cite{Vaitiekenas_PRB2022}\SDEcomment{, and supercurrent reversal \cite{Razmadze_arXiv2022}}.
Concurrent theoretical works demonstrated the possibility of topological superconductivity in these tripartite systems by a combination of a direct induced exchange from the FI into the SM and an indirect one through the SC \Elsa{(present only in overlapping devices)} \cite{Woods_PRB2021, Escribano_PRB2021, Liu_PRB2020, Khindanov_PRB2021, Maiani_PRB2021, Poyhonen_SciPost2021}. A third mechanism whereby electrons tunnel from the SC to the SM through the spin-polarized FI barrier was identified for sufficiently thin FI layers \Elsa{in configurations where the SC and the SM are separated by the FI layer} \cite{Maiani_PRB2021, Langbehn_PRB2021}. \Elsa{We note that in devices where the three materials are in direct contact, a sharp distinction between the three mechanisms is artificial and the overall induced exchange field is due to a combination of all of them.}
In general, fine-tuning from back and side gates was necessary in order to push the SM electron wavefunction close to both the SC and FI layers, maximizing magnetic and superconducting correlations.

In this work we propose a planar SM/FI/SC heterostructure for the creation of a field-free quasi-one dimensional (1D) topological superconductor, Fig.~\ref{Fig1}(a).
In this setup, a thin FI layer is grown between the SC and the SM \footnote{We note that, in principle, a planar SC/SM/FI heterostructure can also exhibit topological properties. However, we do not consider such an arrangement of materials because the growing conditions would lead to a highly disordered heterostructure.}. Due to the band alignment properties between materials, see Fig.~\ref{Fig1}(b), a charge accumulation layer appears at the SM/FI interface~\cite{Liu_PRB2020}, hosting a two-dimensional electron gas (2DEG). The role of the FI layer is two-fold: to induce an exchange field in the 2DEG and the SC, and to act as a spin-polarized barrier for electrons.
In addition, a SC stripe on top of the FI layer defines a quasi-1D geometry where superconductivity is induced. State-of-the-art 2DEG platforms are usually grown on top of rather thick substrates, making it hard to gate them from the bottom. For this reason, we include a top gate in our design, used to manipulate the wavefunction profile in the SM region and drive the system in and out of the topological phase.

To test the properties of this device, we carry out microscopic simulations using a unified numerical approach~\cite{Winkler_PRB2019, Escribano_PRB2021} that describes the electrostatic environment and treats the three different materials on an equal footing. Using specifically an \ch{InAs}/\ch{EuS}/\ch{Al} heterostructure, a robust topological phase appears when the FI thickness is between \SI{\sim 1.5}{nm} and \SI{\sim 3}{nm}, equivalent to 2--5 \ch{EuS} monolayers. It approximately corresponds to the wavefunction penetration length into the FI. In Appendix~\ref{sec:appendix_hexagonalwire} we compare our results with the hexagonal cross-section nanowire geometry, illustrating that our 2D proposal provides larger and more regular topological regions as the external gate is varied, which moreover display larger and harder gaps. We associate this behavior with the stronger vertical confinement achieved in 2DEGs compared to hexagonal nanowires. Therefore, our work establishes 2D ferromagnetic heterostructures as a promising platform for topological superconductivity, opening the possibility of defining complex topological wire structures.

\begin{figure*}
    \centering
    \includegraphics[width=1.9\columnwidth]{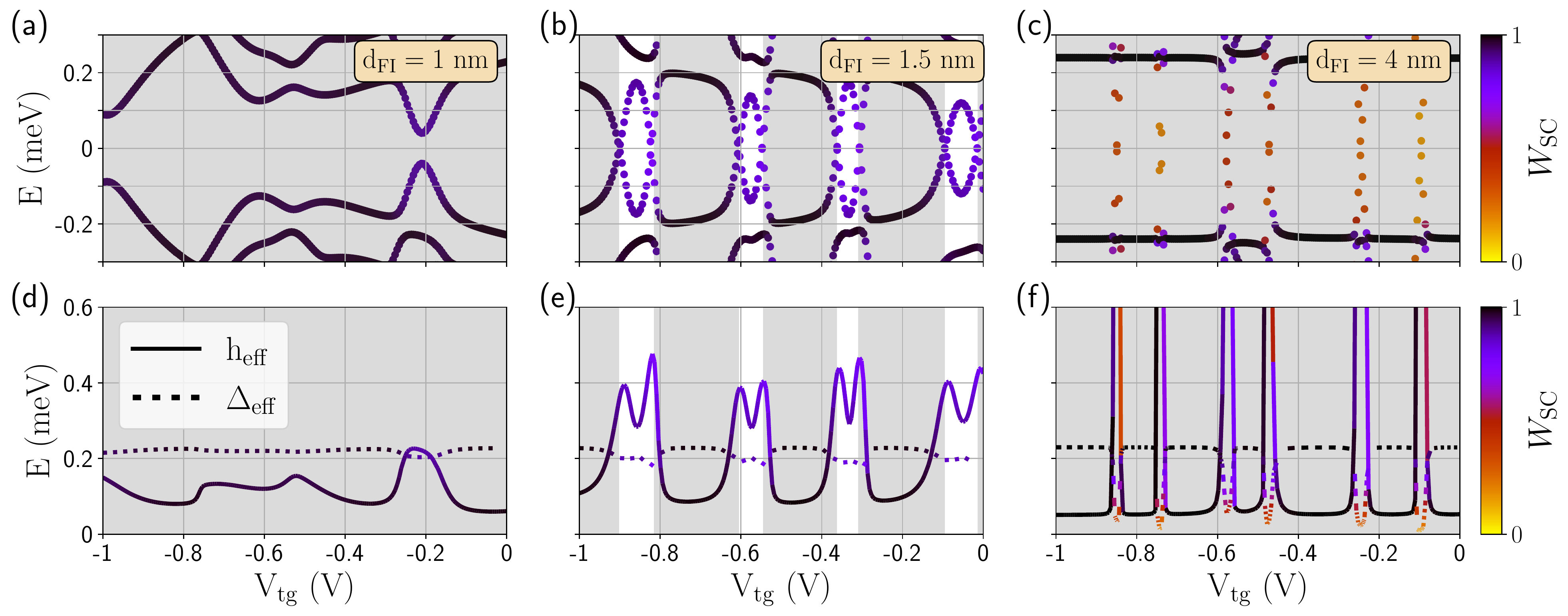}
    \caption{\textbf{Topological phase diagrams for different FI thicknesses.} Top row: energy spectrum at $k_x=0$ as a function of the top-gate voltage $V_{\rm tg}$ for a FI thickness of (a) $d_{\rm FI}=1$~nm, (b)  $d_{\rm FI}=1.5$~nm and (c) $d_{\rm FI}=4$~nm. Colors represent the weight $W_{\rm SC}$ of each state in the superconducting Al layer. Shaded $V_{\rm tg}$ regions are those characterized by a  trivial phase, i.e., $\mathcal{Q}=+1$; while white regions correspond to a topological phase, i.e., $\mathcal{Q}=-1$. Bottom row (d,e,f): effective exchange coupling $h_{\rm eff}$ (solid lines) and superconducting pairing amplitude $\Delta_{\rm eff}$ (dotted lines) for the lowest-energy state in (a), (b), (c), respectively, as given by Eqs.~\eqref{Eq:eff_param1} and~\eqref{Eq:eff_param2}.}
    \label{Fig2}
\end{figure*}

\section{Model and methods}
Following Ref.~\onlinecite{Escribano_PRB2021}, we describe the heterostructure in Fig.~\ref{Fig1}(a) with a Bogoliubov-de Gennes Hamiltonian that includes the conduction band electrons in the three materials. In the Nambu basis $\Psi_{k_x}=(\psi_{k_x \uparrow}, \psi_{k_x \downarrow}, \psi^\dagger_{-k_x \uparrow}, \psi^\dagger_{-k_x \downarrow})$, it is given by
\begin{eqnarray}
H = \left[ \vec{k}^T  \frac{\hbar^2}{2m^*(\vec{r})} \vec{k} + E_{\rm F}(\vec{r}) - e\phi(\vec{r}) +h_x(\vec{r})\sigma_x \right]\tau_z \;\;\;\;\;\;\;\;
\label{eq:hamiltonian}\\ 
 +\frac{1}{2} \left[\vec{\alpha}_R(\vec{r})\cdot \left(\vec{\sigma}\times\vec{k}\right) + \left(\vec{\sigma}\times\vec{k}\right)\cdot \vec{\alpha}_R(\vec{r}) \right]\tau_z + \Delta(\vec{r})\sigma_y\tau_y, \nonumber
\end{eqnarray}
where $\sigma_i$ and $\tau_j$ are the Pauli matrices in spin and Nambu space. We consider a translation invariant system in the $x$-direction. Therefore, the position and momentum operators read as $\vec{r}=(y, z)$ and $\vec{k}=(k_x,-i\partial_y,-i\partial_z)$ in the above Hamiltonian, with $k_x$ being a good quantum number. The model parameters are the effective mass $m^*$, the conduction-band bottom $E_{\rm F}$, the exchange field $h_x$ (non-zero only in the FI), and the superconducting pairing potential $\Delta$ (non-zero only in the SC). Note that $\Delta(\vec{r})$ is real in the above equation. These parameters have a constant value inside each material. For our calculations, we use \ch{InAs} for the SM, \ch{EuS} as FI, and \ch{Al} as SC. The material parameters are given in Table~\ref{Table:parameters} in Appendix~\ref{sec:appendix_model} according to estimations and measurements that can be found in the literature. We also include quenched disorder in the outer surface of the SC, that is characteristic of this kind of heterostructures and beneficial for the superconducting proximity effect~\cite{Winkler_PRB2019, Escribano_PRB2021}. We have found that disorder in the FI (e.g., due to the corrugation of the \ch{EuS}-\ch{Al} interface~\cite{Liu_NL2019}) does not significantly change the energy spectrum (not shown).

We describe the electrostatic interactions in the stacking of Fig.~\ref{Fig1}(a) by solving self-consistently the Schr\"odinger-Poisson equation in the Thomas-Fermi approximation~\cite{Mikkelsen_PRX2018, Winkler_PRB2019, Escribano_PRB2019}.
We take into account the band bending produced at the \ch{InAs}/\ch{EuS} interface~\cite{Liu_AppMat20, Liu_PRB2020}, see Fig.~\ref{Fig1}(b), using a fixed positive surface charge in our simulations. This strong band-bending is crucial as it induces a natural 2DEG at the SM/FI interface, enhancing the topological properties of the device by confining electrons close to the proximitized region \cite{Escribano_PRB2021, Liu_PRB2020}.
Then, a quasi-1D system can be defined by means of an electrostatic lateral confinement.
This is achieved by applying a negative potential to the top gate that depletes the 2DEG everywhere except underneath the grounded SC stripe, which screens the electric field coming from the top gate. This allows controlling the lateral extension (in the $y$-direction) of the SM 1D channels. Moreover, the top gate allows for partial control of the local chemical potential in the effective wire. Our design is independent of the choice of the specific materials as long as they fulfill some requirements: the SM should feature a surface 2DEG, whereas the FI should have a moderate bandgap to allow electron tunneling, and a sufficiently large spin-splitting to induce the topological transition (but small enough not to suppress superconductivity in the SC).

Additional details on the electrostatic problem can be found in Appendix~\ref{sec:appendix_model}. We obtain the self-consistent electrostatic potential~$\phi(\vec{r})$ across the heterostructure along with the Rashba field $\vec{\alpha}_R(\vec r)$, non-zero only in the SM. The Rashba coupling is proportional to the electric field $\vec{\nabla}\phi(\vec{r})$, which is mainly oriented in the $z$-direction, and it is accurately described using the procedure of Ref.~\onlinecite{Escribano_PRR2020} and further discussed in Appendix~\ref{sec:appendix_model}. The spin-orbit field ($\sim \vec{k} \cross \vec{\alpha}_R$) is mainly oriented in the $y$-direction ($\sim \alpha_{R,z} k_x \sigma_y$), with small components in the $x$ and $z$-directions.  We have verified that the electric field in the FI is negligible and, therefore, $\phi(\vec{r})$ is disregarded in that region in Eq.~\eqref{eq:hamiltonian}. 

We describe the FI as a depleted wide-bandgap semiconductor with a spin-split conduction band laying above the Fermi level, as depicted schematically Fig.~\ref{Fig1}(b). 
The topological phase can appear when the FI magnetization is not aligned with the spin-orbit field (which is oriented fundamentally in the $y$-direction in our device), and it is maximized when the magnetization and the spin-orbit field are perpendicular. In this work we assume that the FI exhibits a homogeneous in-plane magnetization along the $x$-direction and negligible stray fields, consistent with the measured easy-axis in thin \ch{EuS}~\cite{Liu_NL2019}. We note that our setup could tolerate in principle an arbitrary misalignment of the exchange field in the $z$-direction since this would still be perpendicular to the spin-orbit term.
This is an advantage with respect to schemes relying on magnetic fields, where relatively small perpendicular magnetic fields to the SC layer suppress superconductivity due to orbital effects.

After the calculation of the electrostatic interactions, we discretize the continuum Hamiltonian in Eq.~\eqref{eq:hamiltonian} following a finite differences scheme with a grid of \SI{0.1}{\nano\m}. We diagonalize the resulting sparse Hamiltonian for different top-gate voltages $V_{\rm tg}$, and longitudinal momenta ~$k_x$ using the routines implemented in Ref.~\onlinecite{Escribano_MNQSP}. From the low-energy eigenstates $\Psi_{k_x}(\vec{r})$ we obtain the topological invariant~\cite{Loring_EurLet10, Zhang_CPB13, Lesser_PRR20, Escribano_PRB2021} and estimate the effective parameters $h_{\rm eff}$ and $\Delta_{\rm eff}$ for the lowest-energy one as
\begin{eqnarray}
    h_{\rm eff}\equiv\left<h_x(\vec{r})\sigma_0\tau_0 \right> \nonumber \\
    = \int \Psi_0^{\dagger}(\vec{r}) h_x(\vec{r}) \sigma_0\tau_0  \Psi_0(\vec{r}) \ \dd\vec{r} = h_0 W_{\rm FI},
\label{Eq:eff_param1}
\end{eqnarray}
\begin{eqnarray}
    \Delta_{\rm eff}\equiv\left<\Delta(\vec{r})\sigma_0\tau_0\right> \nonumber \\
=\int \Psi_0^{\dagger}(\vec{r}) \Delta(\vec{r})\sigma_0\tau_0 \Psi_0(\vec{r})  \ \dd\vec{r} =\Delta_0 W_{\rm SC}, 
\label{Eq:eff_param2}
\end{eqnarray}
\begin{figure*}
    \centering
    \includegraphics[width=1.9\columnwidth]{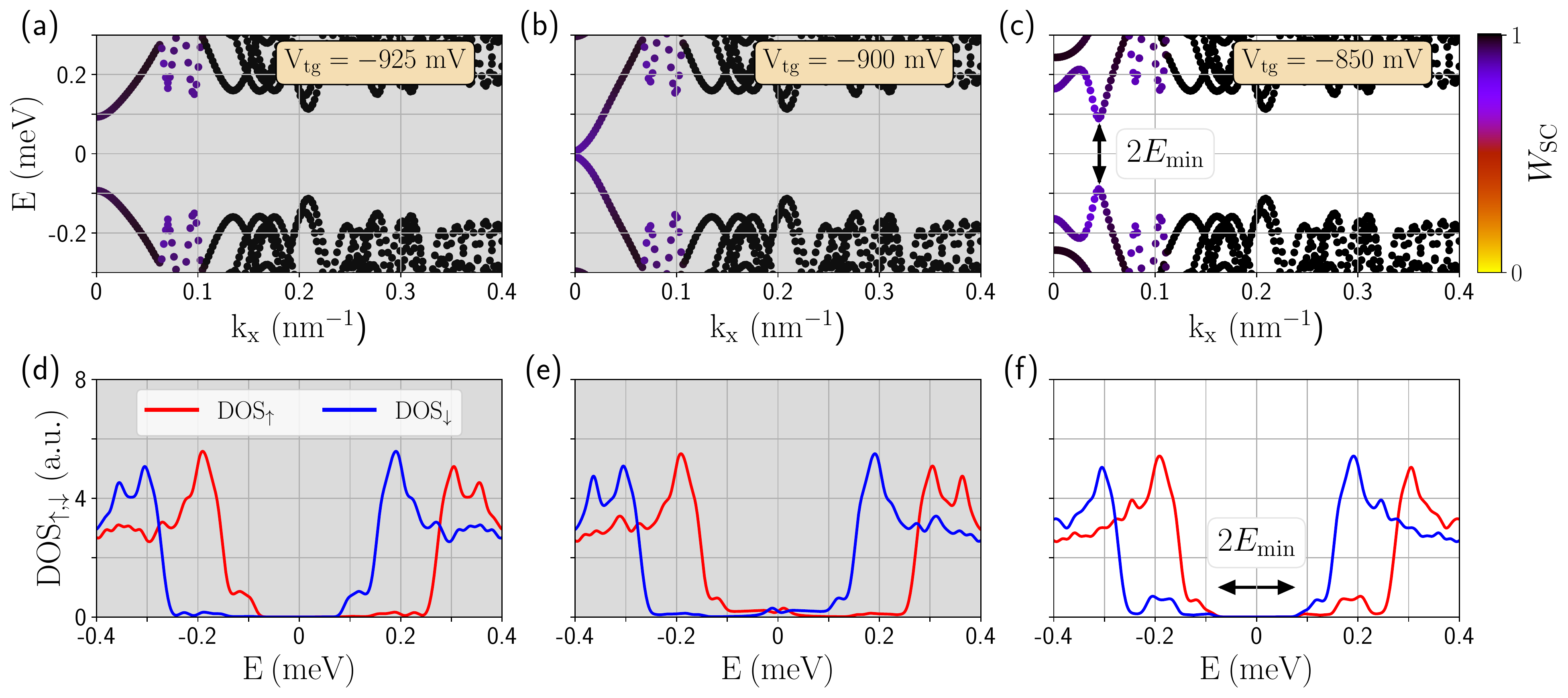}
    \caption{\textbf{Topological phase transition and DOS.} Dispersion relation for a device with \ch{EuS} layer thickness $d_{\rm FI}=1.5$ nm, and for (a) $V_{\rm tg}=-925$ mV (before the topological transition), (b) $V_{\rm tg}=-900$ mV (at the topological transition), and (c) $V_{\rm tg}=-850$ V (in the middle of the topological phase). In (d-f) we show the spin-resolved integrated DOS of the corresponding plot on the top. Only the (c,f) case is topological, with $E_{\rm min}$ being the topological minigap, i.e., the lowest-state energy at $k_x=k_{\rm F}$.}
    \label{Fig3}
\end{figure*}
where $W_{\beta}$ is the weight of the lowest-energy state in the material $\beta=\mathrm{\left\{SC,FI\right\}}$, $\sigma_0$ and $\tau_0$ are the identity matrices in spin and Nambu space, and $h_0$ and $\Delta_0$ are the parent exchange coupling in the FI and the parent superconducting pairing in the SC, respectively. 
The estimation in Eqs.~\eqref{Eq:eff_param1} and \eqref{Eq:eff_param2} is valid for any subgap state ($\left|E_n\right|<\Delta_0$) when the heterostructure thicknesses $d_{\rm SM}\ll \lambda_{\rm SO}$ and $d_{\rm SC}\ll \xi_{\rm SC}$, being $\lambda_{\rm SO}$ the spin-orbit length and $\xi_{\rm SC}$ the superconducting coherence length.
Additional details can be found in Appendix~\ref{sec:appendix_methods}. $h_{\rm{eff}}$ and $\Delta_{\rm eff}$ can be interpreted as the parameters entering in an effective single-band Oreg-Lutchyn Hamiltonian~\cite{Oreg_PRL2010,Lutchyn_PRL2010}
describing the lowest-energy subband.
These quantities, together with the effective chemical potential $\mu_{\rm eff}$, are useful to understand when the system undergoes a topological phase transition, as a large enough exchange field is needed to fulfill the 1D topological criterion, i.e., $\left| h_{\rm eff} \right| \gtrsim\sqrt{\Delta_{\rm eff}^2+\mu_{\rm eff}^2}$~\cite{Oreg_PRL2010}.

\section{Results}
The low-energy wavefunctions decay exponentially in the FI layer on a length scale approximately given by $\xi_\mathrm{FI} = \sqrt{2 E_{\rm F, FI} m^*_\mathrm{FI} / \hbar^2}$, where $E_{\rm F, FI}$ is the conduction band minimum in the FI with respect to the Fermi level. For our materials choice $\xi_\mathrm{FI}\approx\SI{2.3}{\nano\meter}$. As a consequence, the thickness of the FI layer determines the tunneling amplitude between the 2DEG and the SC: thicker FI layers decouple the 2DEG from the SC resulting in a reduction of the superconducting proximity effect, while thinner ones exhibit a reduced induced magnetization in the 2DEG. Hence, there  is an optimal barrier thickness that allows for a sufficiently large induced exchange field and pairing potential in the 2DEG to drive the system into the topological regime.

The topological phase transition of the system occurs at a gap closing and reopening when the lowest energy subband crosses zero energy at the $k_x=0$ high symmetry point. For this reason, in Fig.~\ref{Fig2} we show the energy spectrum of the system at $k_x=0$ as a function of the top-gate voltage for three different values of the FI thickness ($d_{\rm FI}$). The white (gray) background denotes the topological (trivial) phase, determined by the corresponding topological invariant. 

Left panels in Fig.~\ref{Fig2} show the regime where the FI is too thin to induce a topological phase transition. The energy spectrum shows low-energy bands localized mainly in the SC, represented by the black color in Fig.~\ref{Fig2}(a). In this case, superconductivity dominates the properties of the low-energy modes. In Fig.~\ref{Fig2}(b) we show the effective superconducting pairing amplitude (dotted line) and exchange coupling (solid line) calculated using Eqs.~\eqref{Eq:eff_param1} and~\eqref{Eq:eff_param2}. For this thickness, we observe that $h_\mathrm{eff}$ is mostly below $\Delta_\mathrm{eff}$, consistent with the system being in the trivial regime as the topological condition $\left| h_{\rm eff}\right| \gtrsim\sqrt{\Delta_{\rm eff}^2+\mu_{\rm eff}^2}$ cannot be fulfilled.

The situation becomes more favorable for FI layers of intermediate thickness, middle panels in Fig.~\ref{Fig2}. As a function of $V_\mathrm{tg}$, the system shows several topological transitions when consecutive subbands cross zero energy. The topological regions are characterized by a non-trivial topological invariant and are marked by a white background in Figs.~\ref{Fig2}(b) and (e). In these regions, the lowest-energy wavefunction has a significant weight in both the SC and the SM, as illustrated by the purple line color. The topological transition is associated with an increase of~$h_\mathrm{\rm eff}$, overcoming the value of~$\Delta_\mathrm{\rm eff}$, see Fig.~\ref{Fig2}(e). In Appendix \ref{sec:appendix_methods}, we further illustrate that the topological criterion in 1D is fulfilled. The small deviations found are due to the approximated character of the effective parameters. 
We note that, for the optimal range of $d_{\rm FI}$, every subband can be tuned to the topological regime as $V_{\rm{tg}}$ is varied, in contrast to the hexagonal wire case where some subbands do not show a topological crossing, see for instance Ref.~\onlinecite{Escribano_PRB2021} or Appendix~\ref{sec:appendix_hexagonalwire}. This is due to the effective hard-wall confinement of the wavefunction in the thin SM layer in the $z$-direction [see Fig. \ref{Fig1}(a)], which keeps the wavefunction close to the FI/SC layers for every subband. As a consequence, the device shows a regular alternation of trivial and topological regions against $V_{\rm tg}$ with comparable minigaps. 
The topological regions thus occupy a larger area in parameters space compared to the hexagonal wire case, where the appearance of the topological regions is more erratic, since the wavefunction can spread throughout the wide hexagonal section, sometimes avoiding a good proximity effect with the SC/FI layers.

\begin{figure}
    \centering
    \includegraphics[width=1\columnwidth]{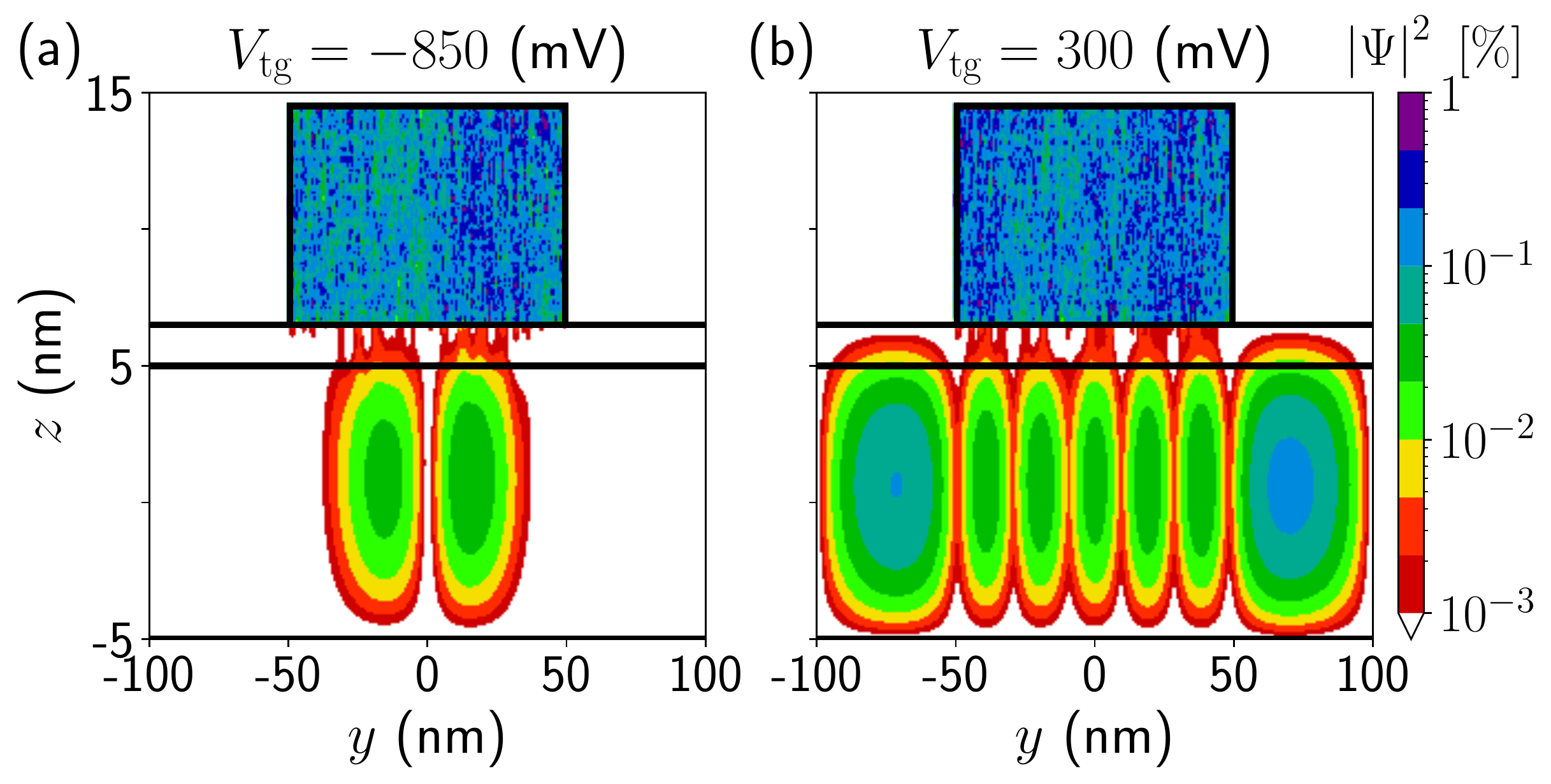}
    \caption{\textbf{Topological and trivial wavefunction profiles.} (a) Transverse probability density at $k_x=0$ for the lowest-energy state in a  topological regime ($V_{\rm tg}=-850$ mV). For comparison, we also show in (b) the case in a topologically trivial regime ($V_{\rm tg}=300$ mV). Parameters are the same as in Fig. \ref{Fig3}, corresponding to $d_{\rm{FI}}=1.5$ nm.}
    \label{Fig3-4}
\end{figure}

The situation of a too-thick FI barrier is illustrated in the right panels of Fig.~\ref{Fig2}. A thick barrier hinders tunneling through the FI, preventing the hybridization of SC and 2DEG states. The reduced hybridization between the two materials can be seen from the shape of the spectrum in Fig.~\ref{Fig2}(c), where the system shows an almost horizontal black line at the SC gap ($E\sim \SI{0.23}{\milli\electronvolt}$) and a series of almost vertical lines (orange dots) crossing the gap. This is also manifested in the abrupt transitions of effective parameters in Fig.~\ref{Fig2}(f). When $\Delta_\mathrm{eff}> h_\mathrm{eff}$ the ground-state wavefunction is localized mostly in the SC and it is essentially independent of the gate voltage, whereas when $\Delta_\mathrm{eff}< h_\mathrm{eff}$ it is localized mostly in the SM. We note that the regions with a large effective exchange field also exhibit a suppressed superconducting pairing, consistent with normal gapless states in the SM.

The properties of a topological superconductor are highly dependent on the value and quality of the topological minigap, which we examine now. In Fig.~\ref{Fig3}, we consider a device with $d_{\rm{FI}}=1.5$ nm as we sweep $V_{\rm{tg}}$. We show the energy subbands versus momentum $k_x$ and the spin-resolved density of states (DOS) in three representative situations: before (left column), at (middle column), and after (right column) the topological transition. Before the transition, Fig.~\ref{Fig3}(a), the heterostructure features a trivial gap and the above-gap states are mostly localized in the SC (black color curves).
The DOS displays a hard gap around zero energy and the characteristic spin-split superconducting coherence peaks, see red and blue curves in Fig.~\ref{Fig3}(d). From this plot we infer that the induced exchange field in the SC is around \SI{100}{\micro\eV} ($\sim50\%$ of the \ch{Al} gap), consistent with the value found in experiments \cite{Hao:PRL91, Strambini_PRM2017, Rouco_PRB19}. A similar peak splitting is found in Figs. \ref{Fig3}(e,f), i.e., it is independent of the value of the gate potential.

At the topological transition, one subband crosses zero energy at $k_x=0$, Fig.~\ref{Fig3}(b). It results in a finite DOS inside the superconducting gap, see Fig.~\ref{Fig3}(e). As we increase $V_{\rm{tg}}$, the superconducting gap reopens in the topological phase, Fig.~\ref{Fig3}(c), accompanied by the onset of Majorana bound states at the ends of a finite-length quasi-1D wire defined by the SC stripe (not shown). The hard gap found in Fig.~\ref{Fig3}(f), $E_{\mathrm{min}}$, has a typical value of tens to a hundred \SI{}{\micro eV}. We associate the large topological gaps found in these devices with the electrostatic confinement in the vertical direction. The thin SM layer, together with the top gate tuned to negative values, makes it possible to concentrate the weight of the wavefunction in the region where superconductivity, magnetism, and spin-orbit coupling coexist. This is signaled by the purple color of the lowest-energy subband in Fig. \ref{Fig3}(c).

The importance of the wavefunction localization is illustrated in Fig.~\ref{Fig3-4}, which shows the lowest-energy wavefunction probability density across the device. In the topological regime, Fig.~\ref{Fig3-4}(a), the ground state wavefunction is concentrated below the SC, maximizing the proximity effects of the SC and FI layers on top. The vertical confinement (in the $z$-direction) is determined by the SM width, $d_{\rm{SM}}$, and the fact that there is an insulating substrate below. The lateral confinement (in the $y$-direction) is achieved by a negative top-gate voltage that depletes the SM everywhere except below the SC. We note that the SM wavefunction penetrates the FI layer all the way to the SC due to its moderate gap and thickness. In the trivial regime shown in Fig.~\ref{Fig3-4}(b), the wavefunction spreads laterally through all the device cross-section (due to a $V_{\rm{tg}}$ value comparable to or larger than the band bending at the SM/FI interface), reducing the proximity effects.

\begin{figure}
    \centering
    \includegraphics[width=1\columnwidth]{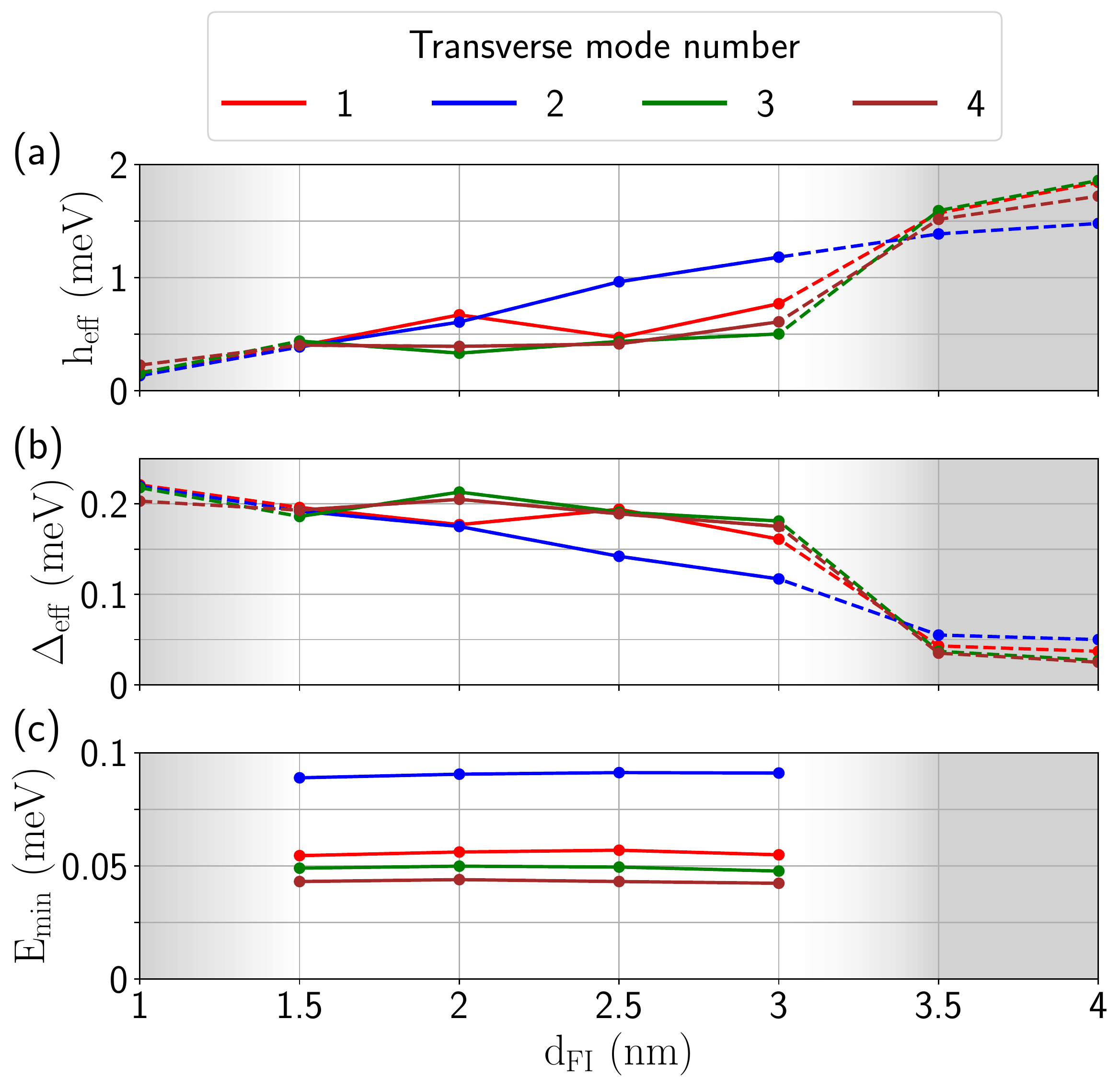}
    \caption{\textbf{Effective parameters as a function of FI thickness.} (a) Effective ferromagnetic exchange coupling $h_{\rm eff}$, (b) effective superconducting pairing amplitude $\Delta_{\rm eff}$, and (c) topological minigap $E_{\rm min}=|E(k_x=k_{F})|$ for the first four occupied transverse subbands (in different colors) versus the EuS thickness $d_{\rm FI}$. We extract these effective parameters when the subband is close to the Fermi level ($E=0$), and therefore different points necessarily correspond to different $V_{\rm tg}$ values. Shaded regions and dashed lines represent that the system is characterized by a topologically trivial phase (and therefore there is no minigap). The suitable FI thickness $d_{\rm FI}$ for topological superconductivity in the 2D stacking device proposed in this work ranges from \SI{\sim 1.5} to  \SI{\sim 3}{nm}.}
    \label{Fig4}
\end{figure}

Finally, we vary the FI thickness to extract the optimal range for topological superconductivity, Fig.~\ref{Fig4}.
The effective exchange coupling is shown in Fig.~\ref{Fig4}(a) and the effective superconducting pairing in Fig.~\ref{Fig4}(b). 
The transverse modes considered (depicted with different colors) are the first four lowest-energy subbands that get populated starting from a depleted SM as we increase $V_{\rm{tg}}$. For each calculated point, we tune $V_{\rm{tg}}$ to the value where the subband is closer to the Fermi level ($E=0$), where $h_{\rm{eff}}$ is maximum, see Fig.~\ref{Fig2}(e,f). Therefore, each point corresponds to a different $V_{\rm{tg}}$ value. We observe that in general $h_{\rm{eff}}$ increases with $d_{\rm{FI}}$ because of the growing weight of the wavefunction inside the FI.
In contrast, the effective superconducting pairing decreases with the FI thickness as the weight of the wavefunction in the SC diminishes.

The topological minigap is shown in Fig.~\ref{Fig4}(c). It is calculated for the value of $V_{\rm{tg}}$ that maximizes $E_{\rm{min}}$ for each subband, i.e., well within the topological region. Depending on the transverse mode, its value ranges from tens to a hundred \SI{}{\micro eV}. Note that we have used the bulk SC gap for the \ch{Al} layer, $\Delta_0=\SI{230}{\micro\electronvolt}$. 
Nevertheless, SCs with larger gaps such as Pb, Nb, Ta, V, or Sn, which can also be grown epitaxially over InAs~\cite{Kanne_NatNAno2021, Bjergfelt_Nanotec19, Khan_AscNano20, Carrad_AdvMat2020}, could help to increase the topological minigap.
Interestingly, for the small SM thickness considered here (\SI{10}{nm}), $E_{\rm min}$ is essentially constant with $d_{\rm{FI}}$ for every transverse mode. This is again a consequence of the vertical confinement that tends to produce regular topological patterns. This regularity gets lost as the SM layer is made thicker, as shown in Appendix~\ref{sec:appendix_extData}. \SDEcomment{In this Appendix we also investigate the role of the SC thickness, finding similar results for thicknesses between \SI{4}{nm} and \SI{12}{nm}.} \Elsa{For a SC with surface disorder, as the one considered here, the induced gap remains essentially unchanged and the main effect of the SC thickness is the renormalization of the SM chemical potential, shifting the picture described in Fig. \ref{Fig2}(b,e) to lower/higher $V_{\rm{tg}}$ values. Thus, the main properties of the topological phase, i.e., regularity, extension, and robustness, do not significantly change, just the value of the top-gate voltage one needs to apply to populate the different subbands.}

Lastly, we have compared our results for the proposed 2D planar heterostructure with a similar stacking in hexagonal nanowires, see Appendix~\ref{sec:appendix_hexagonalwire}. The hexagonal nanowire can also be tuned to the topological regime using an electrostatic gate. However, the topological phase appears for reduced and irregular gate-voltage ranges compared to the planar structure in Fig.~\ref{Fig1}. In addition, the topological gap in hexagonal nanowires is typically soft, exhibiting low-energy trivial states. These states are prone to creating quasiparticle excitations poisoning, undermining coherence in the device and being an obstacle to topological superconductivity. We associate the improved topological properties of the presented 2D stacking with the vertical confinement of the 2DEG wavefunction (see Fig.~\ref{FigSM9} in Appendix~\ref{sec:appendix_hexagonalwire}). \SDEcomment{In contrast, since the quantum well in hexagonal wires is less confined, the wavefunction can spread several nanometers away from the SC/FI interface, giving rise to weaker proximity effects. Moreover, due to the smaller wavefunction localization, the effective parameters are highly dependent on the wavefunction profile (or subband) and, consequently, the phase diagram appears to be more irregular than in the planar device.}

\section{Conclusion}
In this work we have proposed a planar heterostructure for topological superconductivity using a thin ferromagnetic insulator (FI) between a two-dimensional electron gas (2DEG) and a superconductor (SC). The thin FI acts as a spin-filter barrier for electrons tunneling through, inducing a sufficiently large exchange field that gives rise to a topological transition in the tripartite heterostructure. In this geometry, superconducting stripes define quasi-1D wires that can be gated from the top, avoiding bottom gates that might be ineffective due to the rather thick substrates needed to create high-quality semiconducting heterostructures.

For illustration, we have considered an experimentally tested material combination: \ch{InAs} (SM), \ch{EuS} (FI), and \ch{Al} (SC). We have found topological regions for FI thicknesses between $1.5$ and \SI{3}{nm}. Outside this range, the FI is either too thick to allow tunneling between the SC and the SM, or too thin to have a significant influence on the SM electrons. The topological phase features a hard superconducting gap in a range between tens to a $\SI{100}{\micro\electronvolt}$. This constitutes a significant improvement with respect to previous hexagonal nanowire geometries \cite{Vaitiekenas_NatPhys2020, Escribano_PRB2021, Liu_PRB2020}, where these gaps were only possible by fine-tuning side gates to push the wavefunction sufficiently close to the FI/SC layers. We associate this behavior to the vertical confinement of the wavefunction for thin SM layers. Most importantly, this vertical confinement also helps to create a rather regular phase diagram, with topological and trivial phases appearing at controlled values of the top-gate potential. The topological regions produced by the subsequent inverting subbands have moreover a similar $V_{\rm tg}$-range and comparable topological minigaps. Experimentally, this is an advantageous property since it permits to search for the topological phase in a predictable manner rather than by randomly scanning parameters, as it is typically the case with hexagonal nanowires.

\textit{Note added.--} During the preparation of this manuscript, an independent work on a similar subject has been made available as a preprint~\cite{Liu_arXiv2022}. Their results are consistent with the ones of this article~\footnote{In Ref.~\onlinecite{Liu_arXiv2022}, the authors study a similar stacking of materials, although there are some differences with our setup. The SC occupies the whole width of the planar heterostructure (instead of being a SC stripe like in our proposal) and they use periodic boundary conditions in the $y$-direction for its description. They moreover gate the system from the bottom and the exchange field is oriented in the $z$-direction. Despite of this, we agree on the main conclusion that the FI thickness should be of the order of the wavefunction penetration length in order to find topological superconductivity.}.

This version of the article has been accepted for publication, after peer review but is not the Version of Record and does not reflect post-acceptance improvements, or any corrections. The Version of Record is available online at: https://doi.org/10.1038/s41535-022-00489-9.

\section*{Acknowledgments}
We acknowledge insightful discussions with C. Marcus, S. Vaitiek\.{e}nas, L. Galletti, Y. Liu, and C. Schrade. This research was supported by the Spanish Ministry of Economy and Competitiveness through Grants No. PID2020- 11767GB- I00, No. PCI2018-093026, and No. PGC2018-097018- B-I00 (AEI/FEDER, EU), the European Union’s Horizon 2020 research and innovation programme under the FETOPEN Grant Agreement No. 828948 (AndQC) and the María de Maeztu Programme for Units of Excellence in R\&D, Grant No. MDM-2014-0377. We also acknowledge support from the Danish National Research Foundation, the Danish Council for Independent Research  \textbar Natural Sciences, the European Research Council (Grant Agreement No. 856526), the Swedish Research Council, and NanoLund. The research at WIS was supported by the European Union’s Horizon 2020 research and innovation programme
(grant agreement LEGOTOP No. 788715), the DFG (CRC/Transregio 183, EI 519/7-1), the BSF and NSF (2018643), and the ISF
Quantum Science and Technology (2074/19).

\section*{Author contributions}
S. D. E. prepared the numerical codes, performed the simulations, and prepared the figures. E. P. and R. S. S. oversaw the project. S. D. E., A. M., E. P., and R. S. S. wrote the manuscript with contributions from all the authors. All authors contributed to designing the project and to the interpretation of the results.

\section*{Competing interests}
The authors declare no competing interests.

\section*{Data availability}
Data and code are available from the corresponding author upon reasonable request.

\bibliographystyle{apsrev4-2}
\bibliography{bibliography.bib}

\appendix
\section{Details on the model}
\label{sec:appendix_model}

\subsection{Geometry}

In this work we consider a hybrid ferromagnetic planar heterostructure like the one shown in Fig. \ref{Fig1}. It consists of three stacked materials: a semiconductor (SM), a ferromagnetic insulator (FI) and a superconductor (SC).
As host of the 2DEG, we use a \SI{10}{\nano\meter} thick layer of \ch{InAs}. Usually the \ch{InAs} layer is grown on top of an elaborate multilayer semiconductor substrate, used to relax lattice stress and defects. The substrate does not influence the physics of the 2DEG and only affects the electrostatics. Nevertheless, its considerable thickness hinders the use of bottom gates to tune the system properties, except if they are grown together with the substrate, which may increase disorder. We introduce the substrate in the model by taking a \SI{400}{\nano\meter}-thick layer of \ch{In_{0.25}Ga_{0.75}As}.

The \ch{InAs} SM layer has a natural 2DEG appearing at its interface with the FI layer due to the presence of a band bending in that interface. A quasi-1D wire can be defined by applying electrostatic lateral confinement. In our proposal, the lateral confinement is obtained using a grounded superconductor (SC) in the form of a stripe and a top gate. For our calculations, we use a superconducting stripe \SI{100}{\nano\meter} wide and \SI{8}{\nano\meter} thick. A dielectric (\SI{8}{\nano\meter} of \ch{HfO_2}) isolates the top gate from the rest of the system. A negative voltage on the top gate causes the depletion of the 2DEG except below the SC, where an effective 1D wire forms. We include a thin \ch{EuS} layer between the SC and the 2DEG, whose thickness is optimized in this work to improve the topological properties of the device.

We consider that the system is translational invariant along the wire direction, $x$, while the cross-section width in the $y$-direction is \SI{200}{\nano\meter}, see Fig.~\ref{Fig1}(a). The remaining parameters are presented in Table~\ref{Table:parameters}, including typical values for the effective electron mass $(m^*)$, spin-orbit coupling $(\alpha)$, Fermi energy $(E_{\rm F})$, exchange field $(h_{x})$, dielectric constant $(\epsilon)$, and pairing potential $(\Delta)$ for each material.

\begin{table}[t]
\caption{Parameters used for the calculations of this work. Temperature is fixed to \SI{10}{mK} in all our simulations. \label{Table:parameters}}

\renewcommand{\arraystretch}{1.15}
\begin{tabularx}{\columnwidth} {X >{\centering}X >{\centering}X >{\centering\arraybackslash}X}
\hline \hline
\textbf{Material} & \textbf{Parameter} & \textbf{Value} & \textbf{Refs.} \\ 
\hline
InAs & thickness & \SI{10}{nm} & - \\
    & width & \SI{200}{nm} &     \\ \cline{2-4}
    & $m^*$ & 0.023$m_0$  & \onlinecite{Vurgaftman_JAP2001} \\
    & $E_{\rm F}$  & 0  & \onlinecite{Liu_AppMat20}  \\
    & $h_x$ & 0  & - \\
    & $\Delta$ & 0  & - \\
    & $\alpha$ & \multicolumn{2}{c}{(see Sec. \ref{Sec:SOC} for details)} \\ \cline{2-4}
    & $\epsilon_{\rm InAs}$ & 15.5$\epsilon_{0}$ & \onlinecite{Levinshtein:00}  \\
    & $\rho_{\rm surf}$ & $2\cdot 10^{-3}\mathrm{\left(\frac{e}{nm^{3}}\right)}$ & \onlinecite{Thelander_Nanotech2010}, \onlinecite{Reiner_PRX2020} \\ \hline
Al & thickness & \SI{8}{nm} & - \\
    & oxidation thickness & \SI{2}{nm} & \\
    & width & \SI{100}{nm} &     \\ \cline{2-4}
    & $m^*$ & $m_0$  & \onlinecite{Segall_PR1961} \\
    & $E_{\rm F}$  & -\SI{8}{eV}  &  - \\
    & $h_x$ & 0  & - \\
    & $\Delta$ & \SI{0.23}{meV}  & \onlinecite{Chang_NatNano2015} \\
    & $\alpha$ & 0  & \onlinecite{Chiang_EPL2017} \\ \cline{2-4}
    & $V_{\rm SC}$ & \SI{0.4}{eV} & \onlinecite{Liu_PRB2020}  \\ \hline
EuS & thickness & 1--\SI{4}{nm} & - \\ \cline{2-4}
    & $m^*$ & 0.3$m_0$  & \onlinecite{Xavier_PLA1967} \\
    & $E_{\rm F}$  & \SI{0.7}{eV}  & \onlinecite{Alphenaar:09}, \onlinecite{Liu_AppMat20}  \\
    & $h_x$ & \SI{0.1}{eV}  & \onlinecite{Mauger_PR1986}, \onlinecite{Alphenaar:09} \\
    & $\Delta$ & 0  & - \\
    & $\alpha$ & 0  & - \\ \cline{2-4}
    & $\epsilon_{\rm EuS}$ & 10$\epsilon_{0}$ & \onlinecite{Axe_JPCS1969}  \\ \hline
Dielectrics & \ch{In_{0.25}Ga_{0.75}As} thickness & 400 nm & - \\
    & HfO$_2$ thickness & \SI{8}{nm} & \\ \cline{2-4}
    & $\epsilon_{\rm InGaAs}$ & 13.9$\epsilon_{0}$ & \onlinecite{Levinshtein2:00} \\
    & $\epsilon_{\rm HfO_2}$ & 25$\epsilon_{0}$ & \onlinecite{Levinshtein:00} \\
    & $\epsilon_{\rm vacuum}$ & $\epsilon_{0}$ & -  \\
\hline \hline
\end{tabularx}
\end{table}

\subsection{Electrostatic potential}
We compute the electrostatic potential $\phi(\vec{r})$ by solving the Poisson equation across the device cross section,
\begin{equation}
    \vec{\nabla}\cdot\left(\epsilon(\vec{r})\vec{\nabla}\phi(\vec{r})\right)=-\rho(\vec{r})\,,
\label{Eq:Poisson}
\end{equation}
where $\epsilon(\vec{r})$ is the permittivity, which takes a different constant value inside each material. In the right-hand-side, $\rho(\vec{r})$ is the charge density, which includes two terms
\begin{equation}
    \rho(\vec{r})=\rho_{\rm surf}(\vec{r})+\rho_{\rm mobile}(\vec{r}).
\label{Eq:charge}
\end{equation}
The first one, $\rho_{\rm surf}(\vec{r})$, is a one-site-thick layer of positive charge located at the SM/FI interface. It models the band-bending towards this interface that emerges due to the electro-chemical differences between both materials. 
In addition, we impose a Dirichlet boundary condition at the SC surface, $V_{\rm SC}$. Therefore, the SC also contributes to enhancing the SM band bending.
The SC and the FI have different geometries, providing contributions to the band bending with different spatial profiles. Hence, both must be described separately and not with a single parameter. 
The second term in Eq.~\eqref{Eq:charge} is the mobile charge of the conduction band inside the 2DEG. In principle, the problem has to be solved self-consistently by diagonalizing the Hamiltonian of Eq.~\eqref{eq:hamiltonian} together with the Poisson equation~\eqref{Eq:Poisson}. However, both equations decouple under the Thomas-Fermi (TF) approximation for the charge density. This approximation, proven to provide excellent results in heterostructures~\cite{Mikkelsen_PRX2018, Winkler_PRB2019, Escribano_PRB2019}, assumes that the mobile charge is well described by the one of a free 3D electron gas
\begin{eqnarray}
    &\rho_{\rm mobile}&\simeq\rho_{\rm mobile}^{\rm (TF)}= \nonumber \\
    &-\frac{2e\sqrt{2}}{3\pi^2\hbar^3}&\left\{m^*\left|e\phi(\vec{r})-E_{\rm F}\right|f\left[-(e\phi(\vec{r})-E_{\rm F})\right]
    \right\}^{\frac{3}{2}}, \
\end{eqnarray}
where $f(E)$ is the Fermi-Dirac distribution for a given temperature $T$. We do not include the contribution of the valence bands to the mobile charges in the SM for simplicity, since they only play a role for large negative gate potentials. Within this approximation, one still has to solve self-consistently the Poisson equation, as $\rho_{\rm mobile}^{\rm (TF)}$ depends on $\phi(\vec{r})$. However, it does not involve the diagonalization of the Hamiltonian, which is a computationally expensive task. In order to solve the self-consistent scheme, we use an Anderson mixing as explained in Ref.~\onlinecite{Escribano_PRB2021}. A potential at the top gate, together with the boundary conditions explained above, changes the chemical potential in the 2DEG, confining the electrons beneath this region.

\begin{figure}
    \centering
    \includegraphics[width=1\columnwidth]{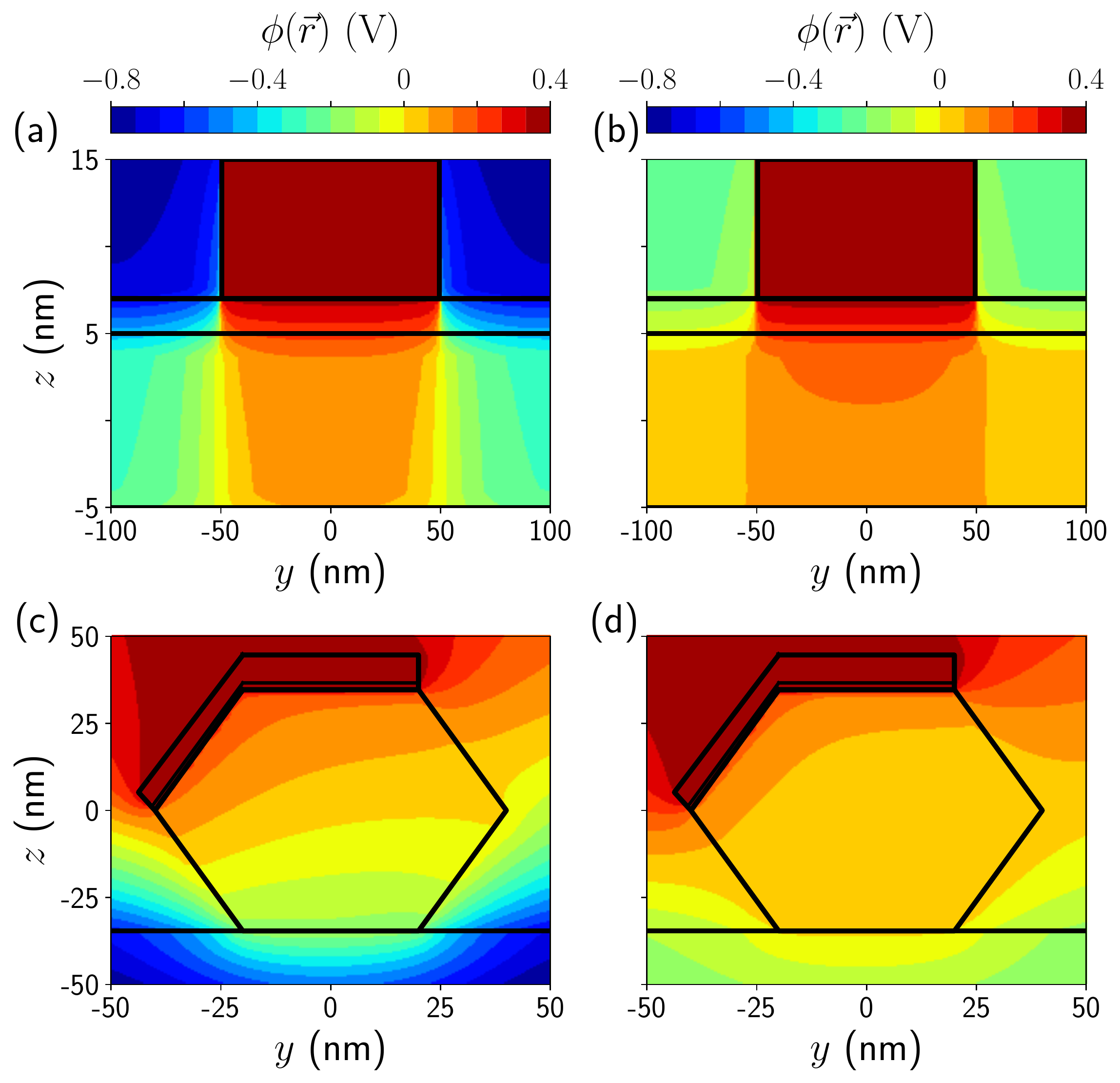}
    \caption{\SDEcomment{ \textbf{Electrostatic potential profiles.} (a, b) Electrostatic potential across the planar device's section for two different top gate potentials: (a) $V_{\rm tg}=-0.8$ V and (b) $V_{\rm tg}=0$ V. In (c, d) we show the same profiles but for the hexagonal wire device, for (c) $V_{\rm bg}=-4$ V and (d) $V_{\rm bg}=0$ V. Black lines represent the interfaces between different materials. See Figs.~\ref{Fig1} and~\ref{FigSM6} for the sketches of the devices.}}
    \label{FigSM10}
\end{figure}

\SDEcomment{To illustrate this, we show in Fig.~\ref{FigSM10}(a) an example of the electrostatic potential profile across the device's section in a highly-depleted regime (i.e., for a large negative $V_{\rm tg}$ potential). Electrons are attracted to positive potential regions, meaning that their wavefunction is localized mainly in these regions. As one can appreciate in Fig.~\ref{FigSM10}(a), when the top gate depletes the system, the electrostatic potential is only positive inside the SM in the region below the SC, due to the screening created by the SC stripe. This effect leads to the confinement effect mentioned previously. On the other hand, when the top gate populates the wire, see Fig.~\ref{FigSM10}(b), the electrostatic potential is positive all across the SM section, and thus, electrons are not only localized beneath the SC. This in turn suppresses the SM/SC coupling and, consequently, this situation is unfavorable to developing a topological superconducting regime inside the SM.}

\subsection{Spin-orbit coupling}
\label{Sec:SOC}
A proper description of the spin-orbit interaction is crucial to predict the robustness of the topological phase~\cite{Klinovaja_PRB12, Winkler_PRB2019}. It arises due to any kind of spatial inversion asymmetry and, therefore, only terms proportional to an odd exponent in $k$ can contribute to this interaction. In this work, we only consider the linear terms in $k$ in the Hamiltonian of Eq.~\eqref{eq:hamiltonian}, as they are the dominant ones, especially in III-V semiconductor compounds. The spin-orbit coupling $\alpha$ mainly depends on the material properties as well as the breaking of the spatial inversion. For Al and EuS, either there is no evidence of a spin-orbit interaction in their bands, or it is negligibly small. However, for III-V semiconductor compounds, like InAs, $\alpha$ is relatively large~\cite{Escribano_PRR2020}. In general, the spin-orbit coupling can be split into two contributions: one arising from a bulk inversion asymmetry $\vec{\alpha}_{\rm D}$, which is called Dresselhaus; and another one which arises from a structural inversion asymmetry $\vec{\alpha}_{R}(\vec{r})$, which is called Rashba. The former depends on the material properties and crystal structure of the compound. While for our choice, zinc-blende (111) InAs, it is negligible, for wurtzite (0001) InAs it can play an important role. These two crystals are the most common ones in the literature as they are easy to grow and possess the smaller lattice mismatch between the SC/FI and the SM~\cite{Korgstrup_NatMat15}. On the other hand, the Rashba component is a spatial dependent function, rather than a constant, since it has to account for the structural inversion asymmetry created by the electrostatic potential. Following the procedure of Ref.~\onlinecite{Escribano_PRR2020}, we describe the SOC as
\begin{equation}
    \vec{\alpha}_{\rm R}(\vec{r})=\frac{eP_{\rm fit}^2}{3}\left[\frac{1}{\Delta_{\rm g}^2}-\frac{1}{(\Delta_{\rm g}+\Delta_{\rm soff})^2}\right]\vec{\nabla}\phi(\vec{r}),
\end{equation}
where $\Delta_{\rm g}$ and $\Delta_{\rm soff}$ are the valence to conduction band gap and split-off gap in the semiconductor. Here, $P_{\rm fit}$ is the Kane coupling (conduction to valence band coupling) corrected to take into account the material and crystal properties of the 2DEG. In our simulations, we model zinc-blende (111) as it has a larger SOC compared to wurtzite structures~\cite{Escribano_PRR2020}. The SOC parameters used in our simulations are given in Table~\ref{Table:parameters_SOC}.

\begin{table}[t]
\caption{Parameters used for the spin-orbit coupling in the InAs, extracted from Ref. \onlinecite{Escribano_PRR2020} and references therein.}
\renewcommand{\arraystretch}{1.15}
\begin{tabularx}{\columnwidth} {X >{\centering}X >{\centering\arraybackslash}X}
\hline \hline
\textbf{Crystal} & \textbf{Parameter} & \textbf{Value} \\ 
\hline
(111) Zinc-blende & $P_{\rm fit}$ & \SI{1300}{meV\cdot nm} \\ 
    & $\Delta_{\rm g}$ & \SI{417}{meV} \\ 
    & $\Delta_{\rm soff}$ & \SI{390}{meV} \\
\hline \hline
\end{tabularx}

\label{Table:parameters_SOC}
\end{table}

\section{Numerical methods, topological invariant and effective parameters}
\label{sec:appendix_methods}
As explained in the main text, we obtain the eigenspectrum of the Hamiltonian of Eq.~\eqref{eq:hamiltonian} by discretizing the space with a regular \SI{0.1}{nm} grid-spacing. We then diagonalize the resulting sparse tight-binding Hamiltonian using the routines implemented in the package of Ref.~\onlinecite{Escribano_MNQSP}. In this way, we obtain the energies $E_n(k_x)$ and their corresponding Nambu-structured eigenstates $\Psi_n(k_x)$, where $n=\left\{1,2,...\right\}$ indexes different transverse modes. This procedure is done for different gate potentials $V_{\rm tg}$ and momenta along the stripe direction, $k_x$. From the eigenpairs we obtain the spin-resolved DOS as
\begin{eqnarray}
    \mathrm{DOS}_{\uparrow\downarrow} (E) = \nonumber \\
    \sum_n \int \dd k_x \Psi_{n}^{\dagger}(k_x) \sigma_{\pm} \Psi_n(k_x)g(E-E_n(k_x)), \\
    g(\omega)=\frac{1}{\sqrt{2\pi k_BT}}e^{-\left(\frac{\omega}{k_BT\sqrt{2}}\right)^2},
\end{eqnarray}
where $\sigma_{\pm}\equiv (\sigma_x\pm i\sigma_y)/2$.

We characterize the system's topology by computing the $\mathbb{Z}_2$ topological invariant
\begin{equation}
    \mathcal{Q}=(-1)^C,
\end{equation}
where $C$ is the Chern number. This can be computed from the low-energy eigenstates through the following expression~\cite{Loring_EurLet10, Zhang_CPB13, Lesser_PRR20, Escribano_PRB2021} 
\begin{equation}
    C=\frac{1}{2\pi}\sum_l \mathrm{Arg}\left\{\lambda_l \right\},
\end{equation}
where $\lambda_l$ are the eigenvalues of the Wilson matrix
\begin{equation}
    \mathcal{W}=\tilde{C}_{-\pi,0}\tilde{C}_{0,\pi}\tilde{C}_{\pi,-\pi},
\end{equation}
being $\tilde{C}_{k_x,k_x'}=\boldsymbol{\Psi}^{\dagger}(k_x)\boldsymbol{\Psi}(k_x')$ the overlapping matrix and $\boldsymbol{\Psi}(k_x)=\left(\Psi_1(k_x), \Psi_2(k_x), ...\right)$ the eigenmatrix containing all the eigenfunctions. As explained in Refs.~\onlinecite{Lesser_PRR20, Escribano_PRB2021}, it is enough to include only the low-energy states in the eigenmatrix $\boldsymbol{\Psi}(k_x)$, as long as the subband that goes topological is included (notice that Majoranas always emerge in the lowest-energy spectrum). This allows to a partial diagonalization of the sparse Hamiltonian in order to compute the topological invariant, reducing the computational cost. 

However, despite its accuracy, the above method cannot explain intuitively why the system transits into the topological phase. To gain some insight, we map the behaviour of each subband of the system to an effective 1D Oreg-Lutchyn model~\cite{Lutchyn_PRL2010,Oreg_PRL2010}, using effective parameters that characterize such band. Our approximation assumes the separability of the eigenfunctions into a product of a purely spatial profile $\psi_n(\vec{r})$, and a position-independent Nambu spinor $\varphi_n$ for the spin and particle-hole subspaces
\begin{equation}
    \Psi_{n}(\vec{r})\simeq \psi_n(\vec{r})\varphi_{n}\,.
\label{Eq:Psi_n}
\end{equation}
This approximation is valid for any subgap state ($\left|E_n\right|<\Delta_0$) when the heterostructure thickness is small compared to the length where the spin and Nambu components change significantly, i.e., the spin-orbit length $\lambda_{\rm SO}$ and the superconducting coherence length $\xi_{\rm SC}$. This is, $d_{\rm{SM}}\ll \lambda_{\rm{SO}}$ and $d_{\rm{SC}}\ll \xi_{\rm{SC}}$. Under this assumption, one can write an effective Oreg-Lutchyn Hamiltonian~\cite{Oreg_PRL2010, Lutchyn_PRL2010} for each transverse subband $n$ as
\begin{equation}
\begin{split}
       H_{\mathrm{eff},n}=\left(\frac{\hbar^2 k_x^2}{2m_{\mathrm{eff},n}}-\mu_{{\rm eff}, n}+h_{{\rm eff}, n}\sigma_x\right)\tau_z \\ +\alpha_{{\rm eff}, n}k_x\sigma_y\tau_z+\Delta_{{\rm eff}, n}\sigma_y\tau_y\,,
    \label{eq:Heff} 
\end{split}
\end{equation}
where the effective parameters are given by 
\begin{equation}
    h_{{\rm eff}, n}\equiv\left<h_x(\vec{r})\sigma_0\tau_0\right>_n=h_0 W_{{\rm FI}, n}\,,
    \label{Eq:h_eff_eq}
\end{equation}
    
\begin{equation}
    \Delta_{{\rm eff}, n}\equiv\left<\Delta(\vec{r})\sigma_0\tau_0\right>_n=\Delta_0 W_{{\rm SC}, n}\,,
\end{equation}

\begin{equation}
\begin{split}
    \mu_{{\rm eff},n}\equiv \left< \bigg( \partial_x\frac{\hbar^2}{2m^*(\vec{r})} \partial_x +\partial_y\frac{\hbar^2}{2m^*(\vec{r})} \partial_y  \right. \\ 
    \left. + E_{\rm F}(\vec{r}) - e\phi(\vec{r}) \bigg) \sigma_0\tau_0 \right>_n\,,
\end{split}
\end{equation}

\begin{equation}
    \alpha_{\rm eff, n}\equiv \left<\alpha_z(\vec{r})\sigma_0\tau_0 \right>_n\,,    
\end{equation}

\begin{equation}
    m_{{\rm eff}, n}^{-1}\equiv \left<\frac{1}{m^*(\vec{r})}\sigma_0\tau_0\right>_n\,,
\end{equation}
being $W_{{\beta}, n}=\int_{\vec{r}\in \beta} \left|\Psi_{n}(\vec{r})\right|^2 \dd\vec{r}$ the weight of the wavefunction in the material $\beta$. Here, $h_0$ and $\Delta_0$ are the parent exchange field in the FI and parent superconducting gap in the SC, correspondingly. In Eq.~\eqref{Eq:h_eff_eq}, we neglect for simplicity additional Zeeman contributions arising from the spin-orbit interaction. 

Notably, the effective Hamiltonian \eqref{eq:Heff} reproduces the spectrum for each band since its eigenvalues $E_n$ (i.e., $H_{\mathrm{eff},n}\varphi_n=E_n\varphi_n$) are the same as the ones of the full Hamiltonian of Eq.~\eqref{eq:hamiltonian}.

\begin{figure}
    \centering
    \includegraphics[width=1\columnwidth]{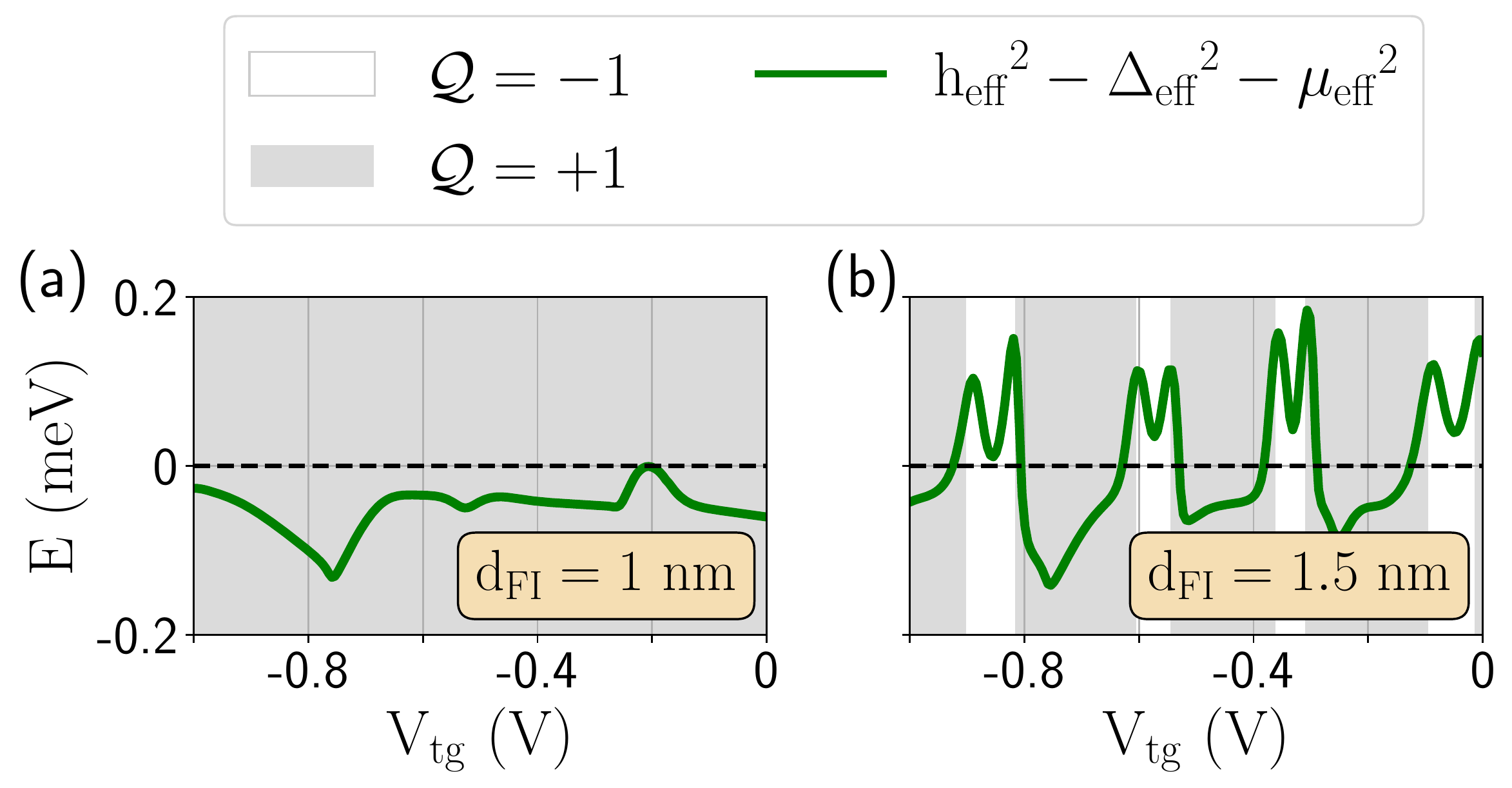}
    \caption{\textbf{Topological invariant vs 1D topological criterion.} Topological criterion comparing results from computing the topological invariant (white/gray background for the topological/trivial phases) and the effective 1D model (green line). We show results as a function of the top-gate voltage $V_{\rm tg}$ and for two different FI thicknesses (a,b). The remaining parameters are the same as in Fig.~\ref{Fig2}.}
    \label{FigSM0}
\end{figure}

Finally, for each subband $n$ that becomes topological when approaching $E=0$ through a gap closing at $k_x=0$, it is possible to verify where the topological criterion 
\begin{equation}
    \left| h_{\rm eff} \right| \gtrsim\sqrt{\mu_{\rm eff}^2+\Delta_{\rm eff}^2}
\end{equation}
is fulfilled as a function of top-gate voltage. We find a good agreement between this criterion and the exact calculation of the topological invariant, as we show in Fig. \ref{FigSM0}. In this figure, we plot the function $f_{\rm top}\equiv h_{\rm eff}^2-\mu_{\rm eff}^2-\Delta_{\rm eff}^2$ vs $V_{\rm{tg}}$ for two different FI thicknesses (a,b). Following the topological criterion, whenever $f_{\rm top}>0$ the system is in the topologically regime. Together with this function, we shade the $V_{\rm tg}$ regions characterized by a positive topological invariant (and therefore, in the trivial phase), calculated using the full spectrum. Note that for the parameters of our device, the total heterostructure thickness is \SI{\sim 20}{nm}, and $\lambda_{\rm{SO}}\sim 100$ nm and $\xi_{\rm{SC}}\sim 100$ \SI{}{nm} for a diffusive SC~\cite{Vaitiekenas_Science2020}. Hence, the condition $d_{\rm{SM}}+d_{\rm{FI}}+d_{\rm{SM}}\ll \lambda_{\rm{SO}},\ \xi_{\rm{SC}}$ is satisfied. We also find the same agreement for the rest of the FI thicknesses (2--\SI{4}{nm}), not shown in Fig. \ref{FigSM0}.

\section{Additional results}
\label{sec:appendix_extData}
\begin{figure}
    \centering
    \includegraphics[width=1\columnwidth]{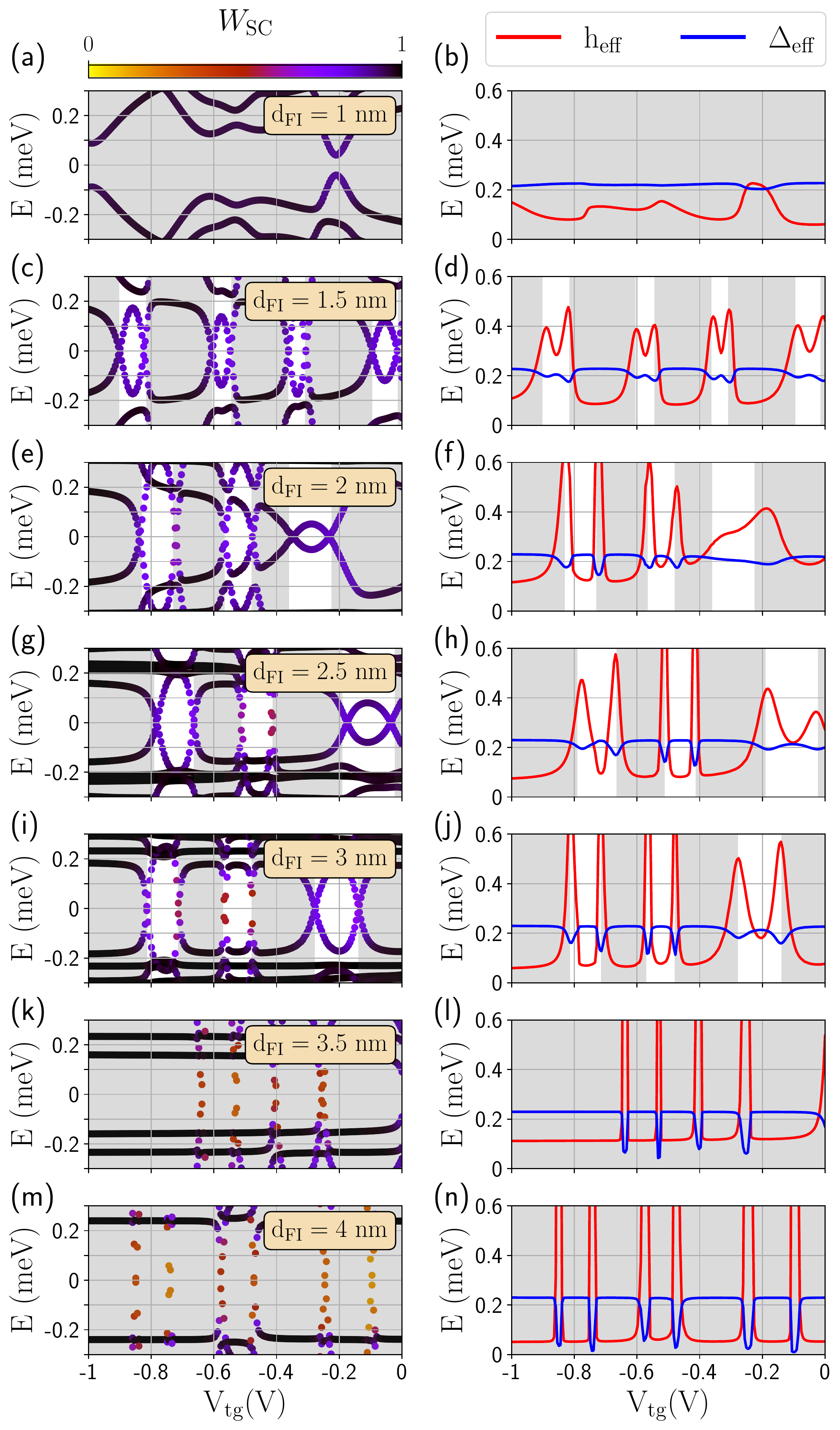}
    \caption{\textbf{Phase diagrams for different FI layer thicknesses.} Energy spectrum at $k_x=0$ versus the top-gate voltage $V_{\rm tg}$ (left panels) for different thicknesses of the \ch{EuS} layer, $d_{\rm FI}$ (different rows). Colors represent the weight of each state on the superconducting \ch{Al} layer, $W_{\rm SC}$. Shaded background $V_{\rm tg}$ regions are characterized by a trivial topological phase, i.e., $\mathcal{Q}=+1$; while white background ones correspond to the non-trivial phase, i.e., $\mathcal{Q}=-1$. Right panels: effective exchange coupling $h_{\rm eff}$ and superconducting pairing amplitude $\Delta_{\rm eff}$ for the lowest energy state on the left.}
    \label{FigSM1}
\end{figure}

In this section, we show complementary results to the ones in the main text, including different values for the \ch{EuS}, \ch{Al} and \ch{InAs} thickness. In Fig.~\ref{FigSM1} we show the low-energy bands (left panels) and the effective exchange field and superconducting gap (right panels), as done in Fig.~\ref{Fig2} in the main text but for more values of the \ch{EuS} thickness. These results were used for the ones presented in Fig. \ref{Fig4} of the main text. From these, it becomes clear that the system can be tuned to the topological regime for $d_{\rm FI}$ between $1.5$ and \SI{3}{nm}. Above these thicknesses, the lowest-energy wavefunction are confined either in the SM or the SC, illustrated by values $W_{\rm SC}<0.5$ or $W_{\rm SC}\sim1$.

\begin{figure}[h]
    \centering
    \includegraphics[width=1\columnwidth]{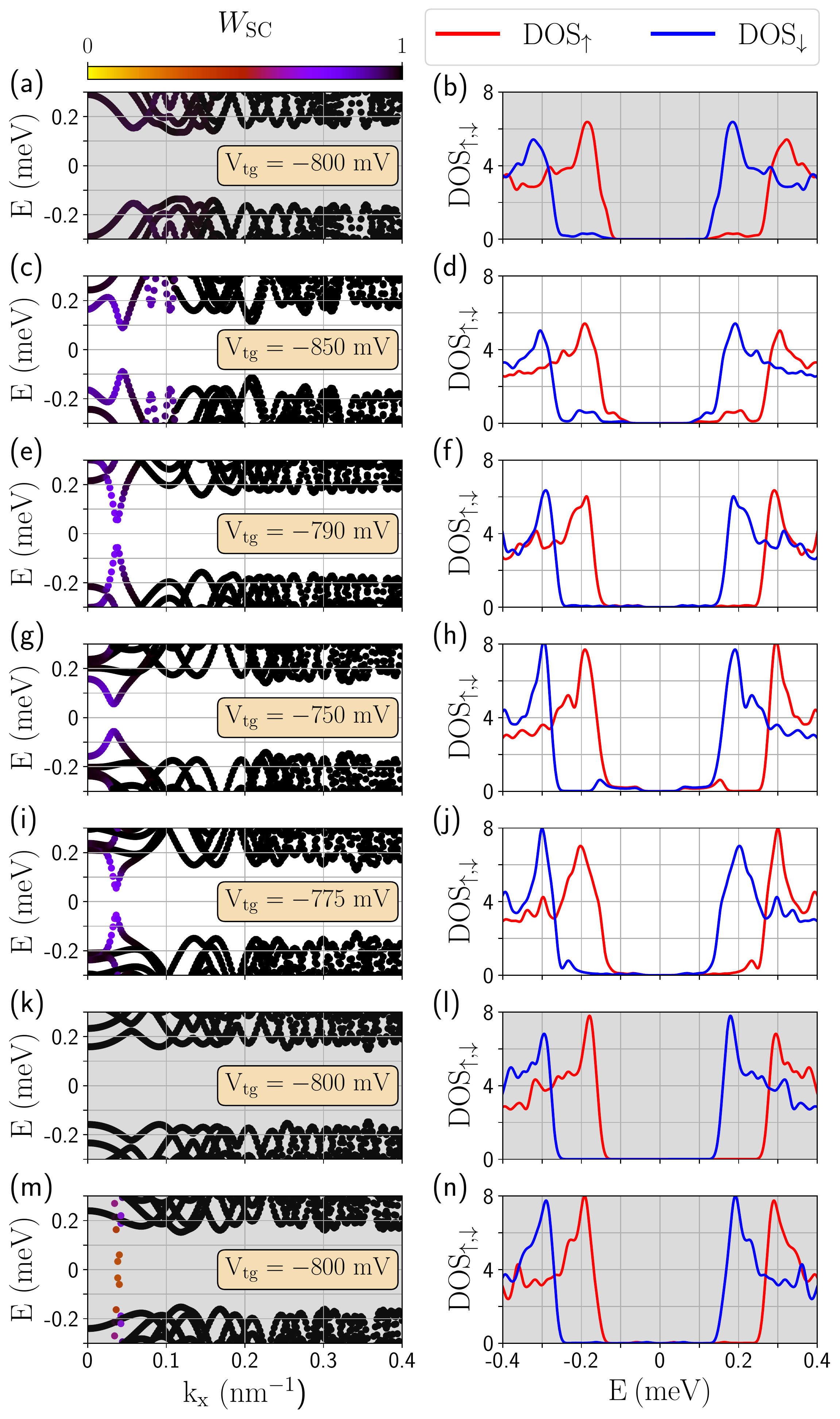}
    \caption{\textbf{DOS for different FI layer thicknesses.} Dispersion relation for (a) $d_{\rm FI}=1$ nm and $V_{\rm tg}=-800$ mV, (c) $d_{\rm FI}=1.5$ nm and $V_{\rm tg}=-850$ mV, (e) $d_{\rm FI}=2$ nm and $V_{\rm tg}=-700$ mV, (g) $d_{\rm FI}=2.5$ nm and $V_{\rm tg}=-750$ mV, (i) $d_{\rm FI}=3$ nm and $V_{\rm tg}=-775$ mV, (k) $d_{\rm FI}=3.5$ nm and $V_{\rm tg}=-800$ mV and (m) $d_{\rm FI}=4$ nm, and $V_{\rm tg}=-800$ mV. In (b,d,f,h,j,l,n) we show the integrated spin-resolved DOS (in a.u.) of the corresponding plot on the left. Notice the hard gap, i.e., the absence of states below $E_{\rm min}$.}
    \label{FigSM2}
\end{figure}

For completeness, we also show in Fig.~\ref{FigSM2} the energy spectrum versus momentum for the potential gate at which the first subband develops a non-trivial topology, when present. In all the represented cases, the system exhibits a hard gap in the topological regime. The minigap data of Fig.~\ref{Fig4} for the first subband is extracted from here.

Several parameters are either unknown or sample-dependent. For this reason, we have performed additional calculations to check the robustness of our conclusions against changes of these parameters. One of them is the band-bending of the 2DEG towards the FI interface. This band-bending controls the ungated doping of the wire and the strength of the hybridization between the 2DEG states with the FI and the SC. Hence, its precise value may affect the topological properties of the wire as well as the optimal FI thickness to have a topological phase.

\begin{figure}
    \centering
    \includegraphics[width=1\columnwidth]{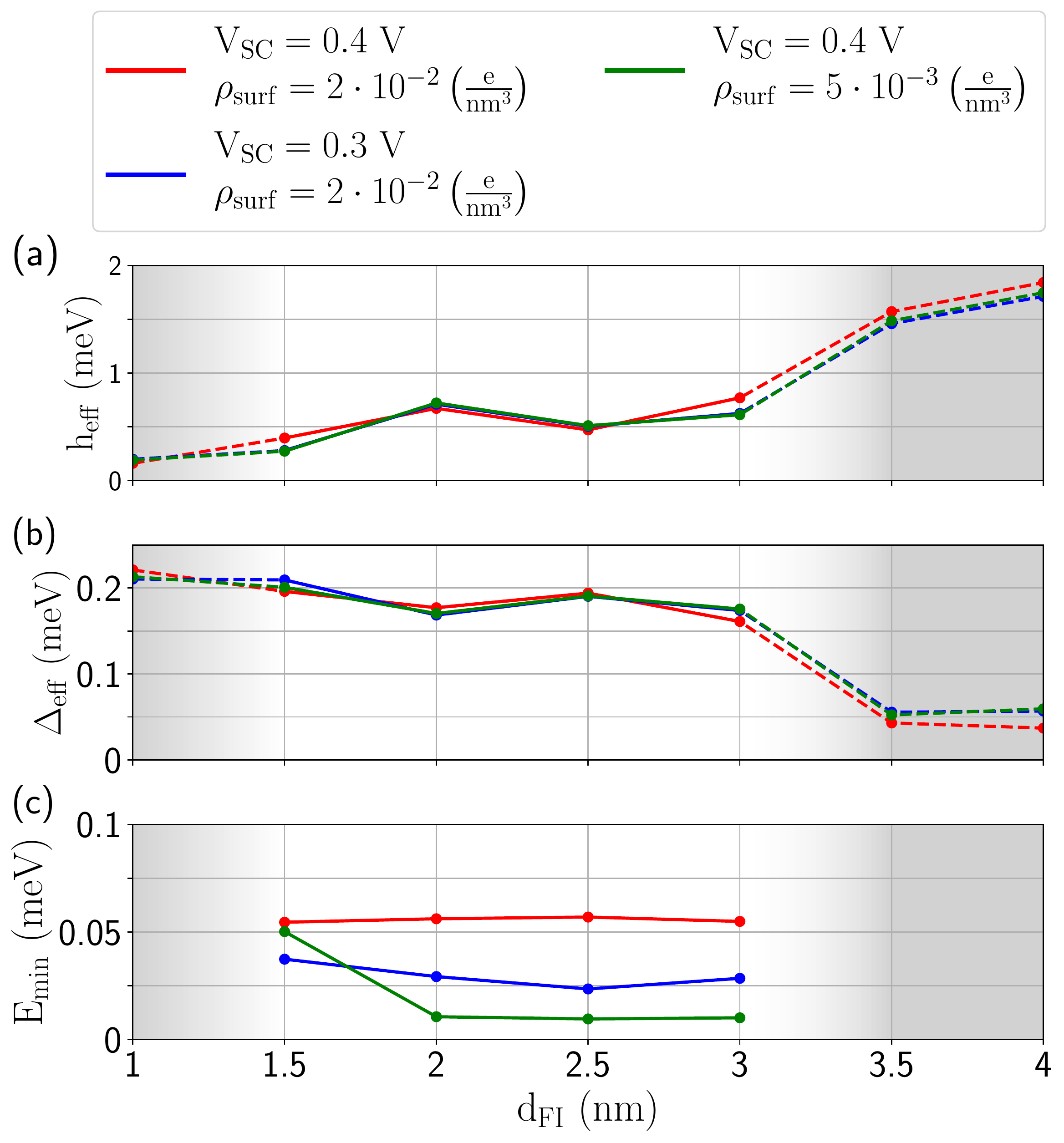}
    \caption{\textbf{Effective parameters for different electrostatic parameters.} Effective exchange coupling $h_{\rm eff}$ (a), superconducting pairing amplitude $\Delta_{\rm eff}$  (b), and minigap $E_{\rm min}=|E(k_x=k_{F})|$ (c) for the first transverse subbands versus the \ch{EuS} thickness $d_{\rm FI}$. Different curves correspond to different choices of electrostatic parameters (see legend). Shaded and dashed regions represent that the system is characterized by a topologically trivial phase (and therefore the minigap is undefined).}
    \label{FigSM3}
\end{figure}

The band-bending depends on two parameters: the surface charge at the 2DEG/FI interface $\rho_{\rm surf}$ and the potential at the SC boundaries $V_{\rm SC}$. Both lead to a charge accumulation close to the 2DEG/FI interface. In our system, $\rho_{\rm surf}$ is homogeneous across the interface while $V_{\rm SC}$ concentrates charges below the SC region. In Fig.~\ref{FigSM3} we show the effective exchange field (a), the effective superconducting gap (b), and the topological gap (c) for the first subband as a function of the \ch{EuS} thickness. Different curves correspond a different band-bending profiles, dependent on $\rho_{\rm surf}$ and $V_{\rm SC}$. In red, we show the case in the main text for comparison. The blue line correspond to a smaller value of $V_{\rm SC}$, and the green one for a smaller value of $\rho_{\rm surf}$, decreasing the charge accumulation with respect to the result shown in the main text. The three cases are qualitatively similar, indicating that a smaller band-bending does not affect significantly the effective parameters of the lowest subband. Therefore,  the optimal \ch{EuS} thickness remains the same as the one found in the main text. 

However, the number of transverse subbands that develop a topological phase reduces with respect to the case shown in the main text (not shown here). Actually, if one decreases more these parameters, for example $V_{\rm SC}\le 0.2$ V and/or $\rho_{\rm surf}\le 2\cdot 10^{-3} \mathrm{\left(\frac{e}{nm^3}\right)}$, it is not possible to find a topological state for any $V_{\rm tg}$ or $d_{\rm FI}$ (not shown). The reason is that the band-bending dramatically changes the doping of the wire. If the initial doping of the 2DEG is too small, a positive potential must be used to effectively dope it. Due to the SC screening, only the regions away from the wire will be populated, leading to poor proximity effects and no topological states. This problem can be fixed using a back gate tuning the doping of the SM. In contrast, if the band-bending is large enough, a negative top gate potential deplete the 2DEG. This in turn confines the wavefunction below the SC, enhancing the proximity effects and the gap in the topological regime, Fig.~\ref{FigSM3}(c).

\begin{figure}
    \centering
    \includegraphics[width=1\columnwidth]{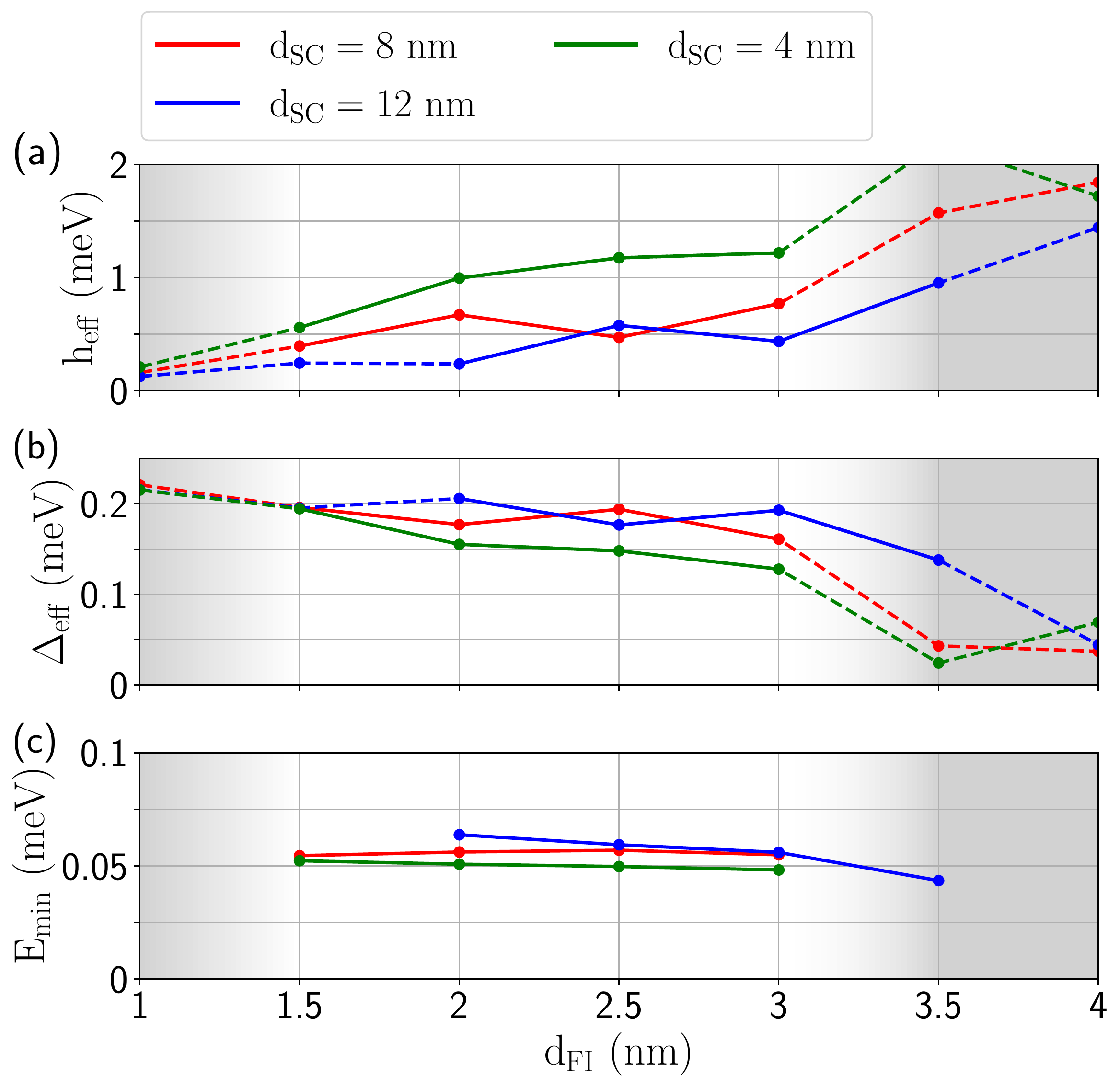}
    \caption{Same as Fig.~\ref{FigSM3} but different curves correspond to different SC thicknesses.}
    \label{FigSM4}
\end{figure}

Apart from these electrostatic parameters, the thickness and width of the different layers can be tuned to optimize topological properties. In particular, we simulate different values for the SC and the SM thickness, which are experimentally controllable.  In Fig.~\ref{FigSM4}, we show the effective exchange field, pairing potential, and topological gap for different  SC thickness. The red one corresponds to the one studied in the main text ($d_{\rm SC}=8$ nm) and the blue and the green ones correspond to a thicker ($d_{\rm SC}=12$ nm) and thinner ones ($d_{\rm SC}=4$ nm). The three curves seem to have a similar optimal $d_{\rm FI}$ range between $1.5$ and \SI{3}{nm}. We note, however, that the topological window shifts to larger $d_{\rm FI}$ value for the thicker SC considered. In addition, a thicker SC exhibits a reduced effective exchange field and larger superconducting pairing amplitude, as shown in Fig.~\ref{FigSM4}(a,b). This leads to a larger topological minigap compared to the thinner SC case, Fig.~\ref{FigSM4}(c). The reason is  the increased electron confinement inside the SC when increasing $d_{\rm SC}$.

\begin{figure}
    \centering
    \includegraphics[width=1\columnwidth]{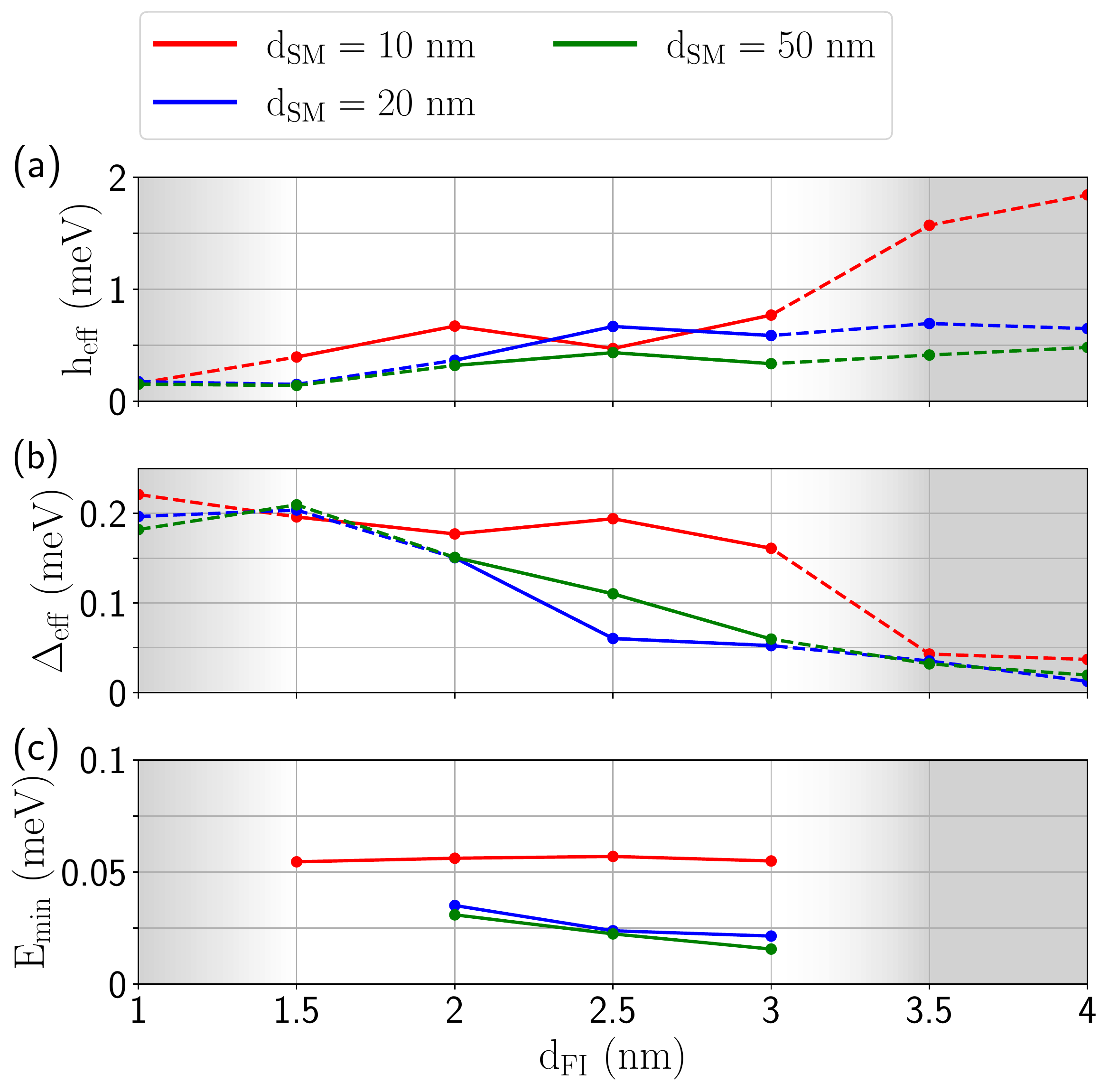}
    \caption{Same as Fig.~\ref{FigSM3} but different curves correspond to different SM thicknesses.}
    \label{FigSM5}
\end{figure}

Finally, in Fig. \ref{FigSM5} we analyze the effect of the 2DEG thickness.
Increasing this thickness enlarges the wavefunction delocalization across the section of the SM, diminishing the electron hybridization between the FI and SC layers. Therefore, the effective exchange field and superconducting gap is reduced as the SM thickness is increased, as illustrated by Figs.~\ref{FigSM5}(a,b). Hence, the range of FI thicknesses where the systems shows topological properties is reduced. Moreover, their topological gap is smaller, Fig.~\ref{FigSM5}(c). This is in agreement with our observations of the same stack in hexagonal nanowires (see Appendix~\ref{sec:appendix_hexagonalwire}), which exhibit worse topological properties due to the same wavefunction delocalization. This illustrates the crucial role of electron confinement for creating topological superconductivity.

\section{Hexagonal nanowire geometry}
\label{sec:appendix_hexagonalwire}

The SM/FI/SC stack can also be grown in a vapor-liquid-solid (VLS) hexagonal nanowire geometry. Recent experiments have shown that it is possible to grow an epitaxially layer of \ch{EuS} on selected facets of \ch{InAs} nanowires, followed by epitaxial \ch{Al} on top, partially or totally overlapping with \ch{EuS} \cite{Vaitiekenas_NatPhys2020, Vaitiekenas_PRB2022}. In this section, we analyze the spectrum and topological properties of such a structure to ascertain whether this platform would be better than the planar heterostructure presented in the main text. 

\begin{figure}
    \centering
    \includegraphics[width=0.7\columnwidth]{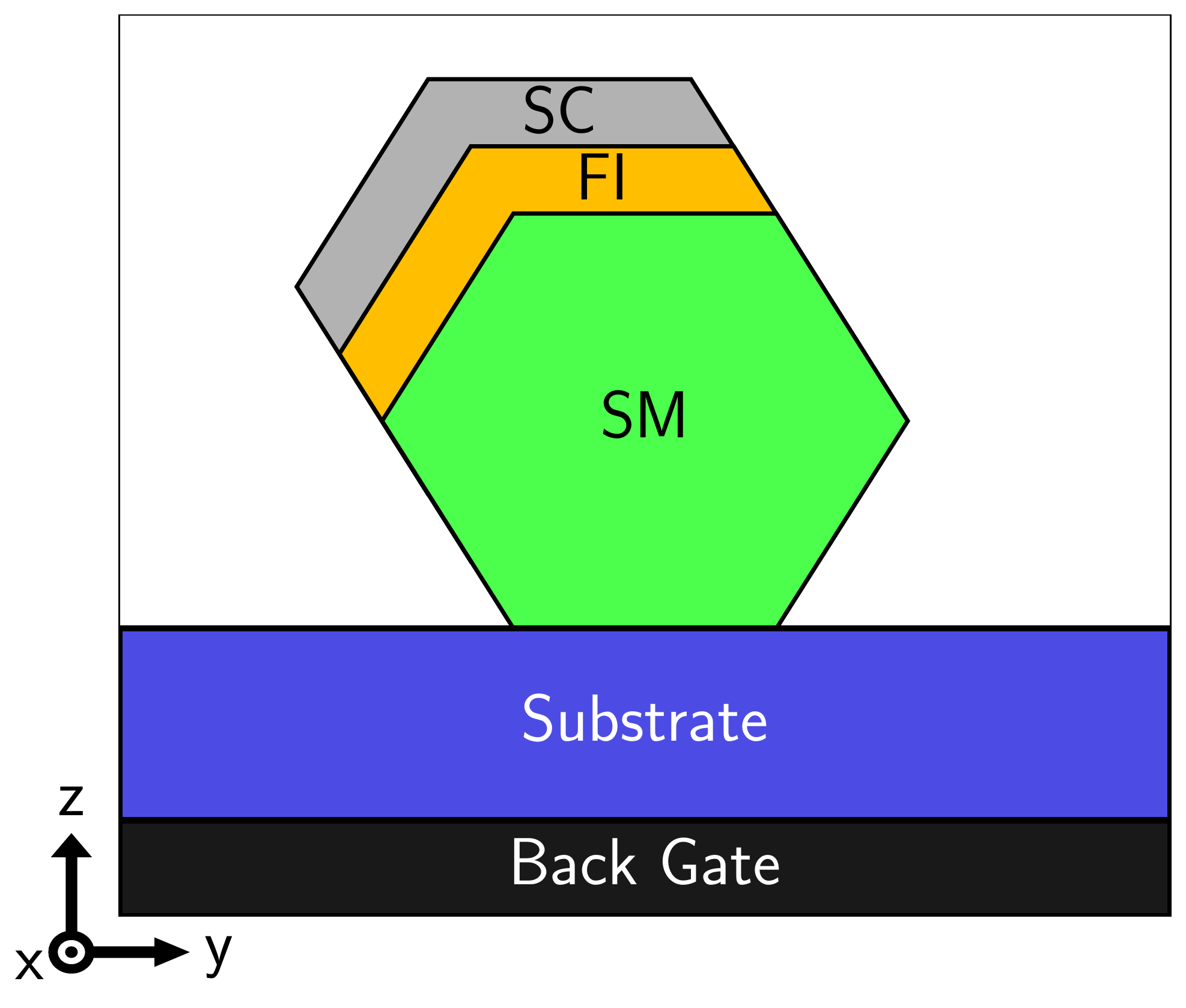}
    \caption{\textbf{Hybrid nanowire heterostructure.} Sketch of the hexagonal nanowire geometry. An hexagonal SM nanowire is partially covered by a FI layer. On top of the FI, a grounded SC layer is included. The nanowire is gated from below using a back-gate isolated from the wire by a \SI{200}{nm} thick \ch{SiO_2} dielectric (blue).}
    \label{FigSM6}
\end{figure}

We describe the system using the Hamiltonian of Eq.~\eqref{eq:hamiltonian} in the main text and the geometry shown in Fig.~\ref{FigSM6}. The hexagonal \ch{InAs} nanowire (green) of \SI{80}{nm} width is covered over two facets by a thin \ch{EuS} (FI) layer (yellow). The outer facets of the \ch{EuS} layer are covered in turn by an \SI{8}{nm} thick \ch{Al} layer (grey). The wire is deposited on top of a \SI{200}{nm} thick \ch{SiO_2} dielectric (blue), and gated from below through a back gate (black). The parameters that we use for the simulations are the same as the ones given in Table~\ref{Table:parameters}, except for these geometrical ones (we also use $\epsilon_{\rm SiO_2}=3.9\epsilon_0$ for the substrate). We highlight that \SI{80}{nm} is the typical diameter for these nanowires, much larger than the SM thickness in the 2DEG-based devices analyzed in the main text.

\SDEcomment{The electrostatic potential profile across the hexagonal nanowire's section is shown in Fig.~\ref{FigSM10}(c,d), together with the planar device's profiles (a,b) for comparison. In Fig.~\ref{FigSM10}(c) we show the case where the back gate strongly depletes the wire. The electrostatic potential is larger close to the FI/SC interface due to the band-bending present there. However, the electrostatic potential is positive in the entire upper-half of the wire, i.e., up to $\sim$\SI{30}{nm} away from the SM/FI interface. Hence, the electron wavefunction will spread all across that region [positive $\phi(\vec{r})$], and therefore its localization close to the SC will be smaller compared to the planar device. This necessarily leads to a worse hybridization with both the FI and the SC. For completeness, we show in Fig.~\ref{FigSM10}(d) the case where the back gate fills the wire (i.e., positive $V_{\rm bg}$). In this case the electrostatic potential is nearly homogeneous across the wire's section and thus, the hybridization between the SM and the FI and SC will be almost completely suppressed.}

\begin{figure*}[ht]
    \centering
    \begin{minipage}{0.48\linewidth}
        \centering
    \includegraphics[width=0.994\columnwidth]{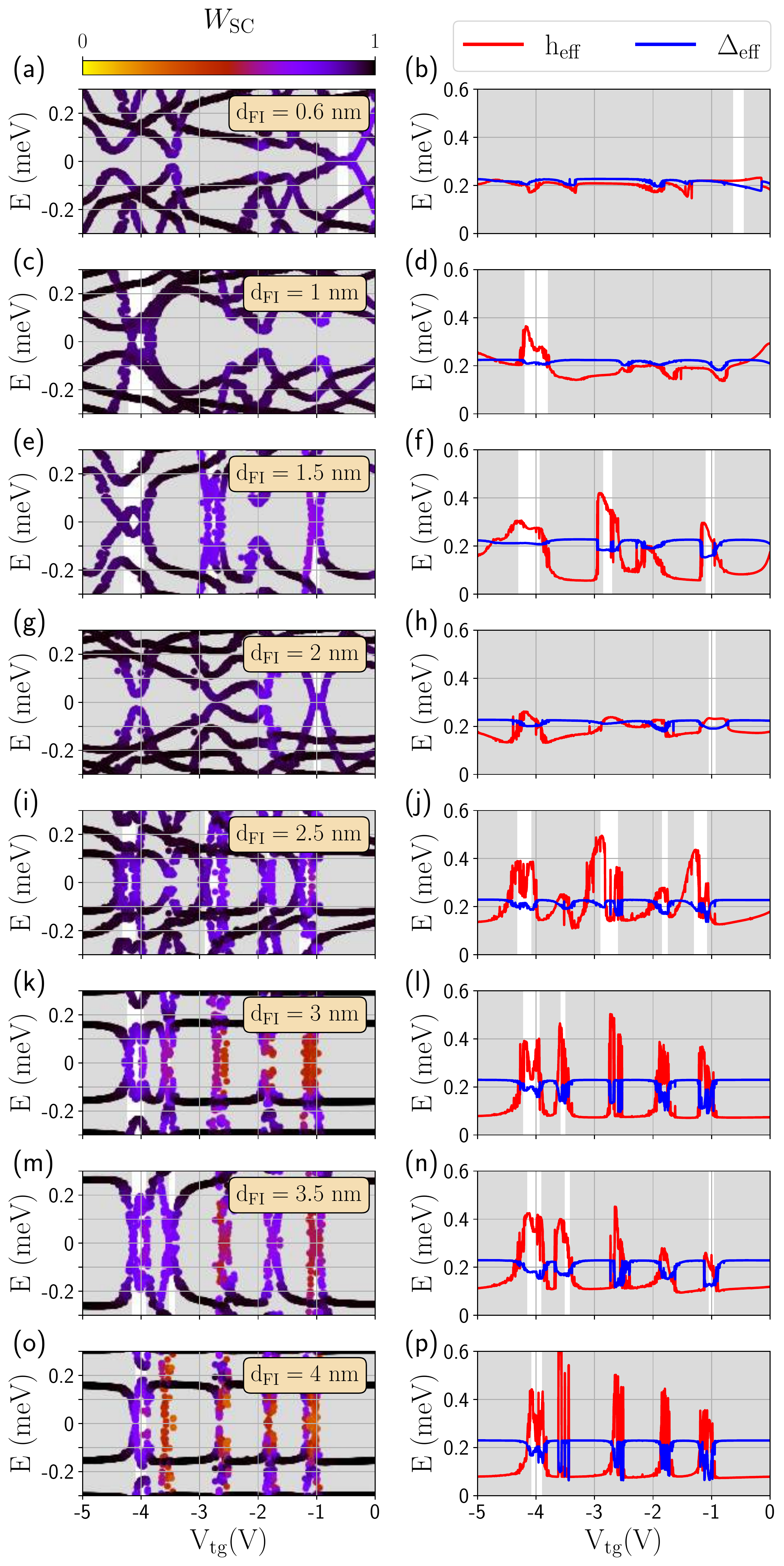}
    \caption{\textbf{Phase diagrams for different FI layer thicknesses for the nanowire device.} Energy spectrum at $k_x=0$ versus the back gate voltage $V_{\rm bg}$ (left panels) for different thicknesses of the EuS layer $d_{\rm EuS}$ (different rows) for the hexagonal wire device (see sketch of the device in Fig.~\ref{FigSM6}). Colors represent the weight of each state on the superconducting Al layer $W_{\rm SC}$. Shaded $V_{\rm tg}$ regions are those characterized by a trivial topological phase, i.e., $\mathcal{Q}=+1$; while the lighter ones correspond to non-trivial ones, i.e., $\mathcal{Q}=-1$. Right panels: effective exchange field $h_{\rm eff}$ and superconducting gap $\Delta_{\rm eff}$ for the lowest energy state on the left.}
    \label{FigSM7}
    \end{minipage}\hfill
    \begin{minipage}{0.48\linewidth}
\centering
    \includegraphics[width=0.994\columnwidth]{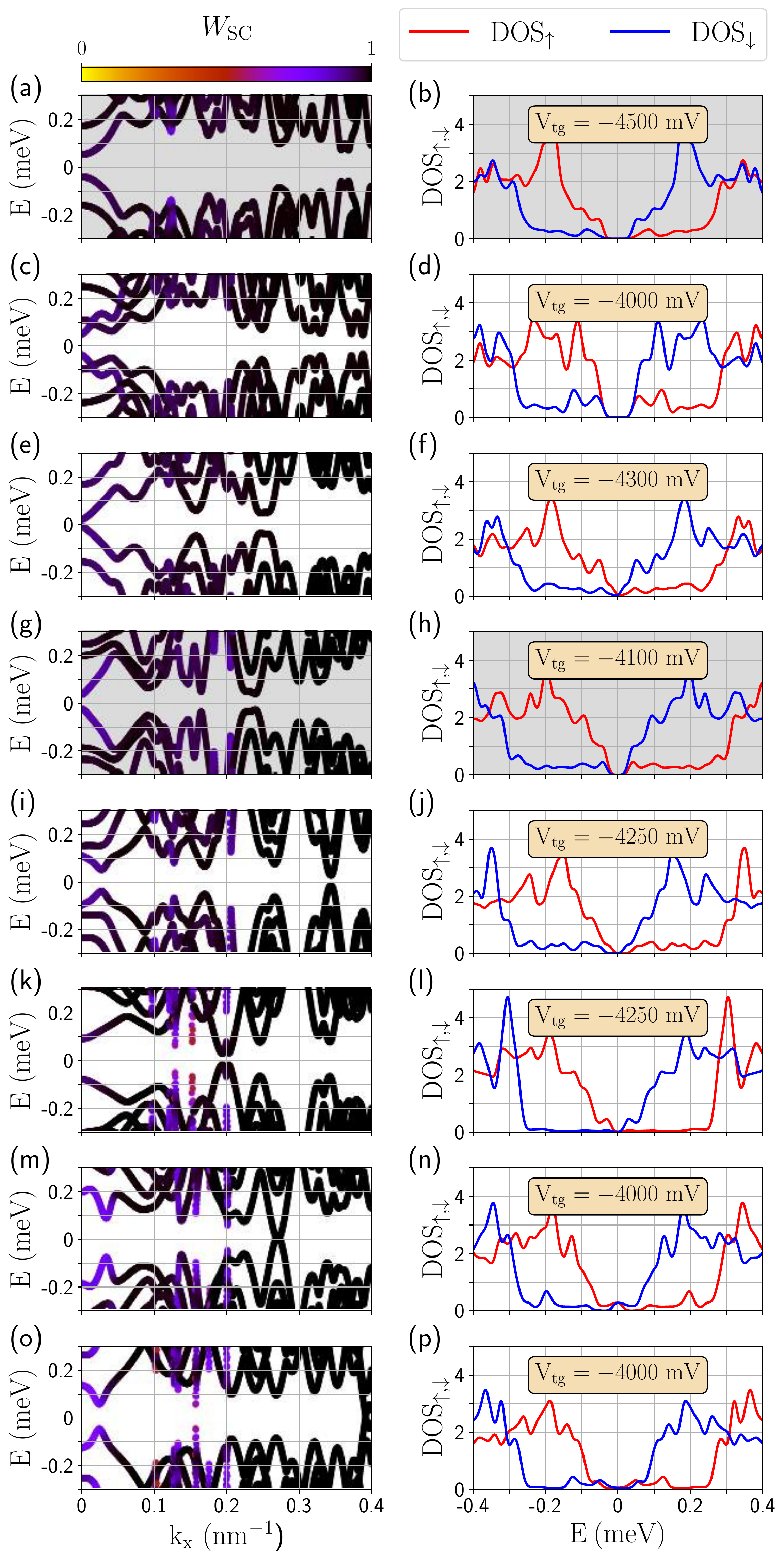}
    \caption{\textbf{DOS for different FI layer thicknesses for the nanowire device.} Left panels, dispersion relation for an hexagonal nanowire for (a) $d_{\rm FI}=0.6$ nm and $V_{\rm bg}=-4.5$ V, (c) $d_{\rm FI}=1$ nm and $V_{\rm bg}=-4$ V, (e) $d_{\rm FI}=1.5$ nm and $V_{\rm bg}=-4.3$ V, (g) $d_{\rm FI}=2$ nm and $V_{\rm bg}=-4.1$ V, (i) $d_{\rm FI}=2.5$ nm and $V_{\rm bg}=-4.25$ V, (k) $d_{\rm FI}=3$ nm and $V_{\rm bg}=-4.25$ V, (m) $d_{\rm FI}=3.5$ nm and $V_{\rm bg}=-4$ V, and (o) $d_{\rm FI}=4$ nm and $V_{\rm bg}=-4$ V. In the right panels, we show the integrated spin-resolved DOS of the corresponding plot on the left.}
    \label{FigSM8}
    \end{minipage}
\end{figure*}

The energy spectrum at $k_x=0$ \SDEcomment{vs the back gate potential} is shown in the left panels of Fig.~\ref{FigSM7} for different thicknesses of the FI layer. We show the topological (trivial) phase as a white (gray) background. As shown in the figure, it is possible to tune the system in the topological regime for a wider thicknesses of the FI barrier compared to the planar structure shown in the main text. However, these phases are narrower and appear in a less regular way than the case in the main text. This is related to the fact that some bands cannot be tuned to the topological regime as they cannot be confined to the interesting spatial region where superconductivity and exchange field coexists. Therefore, the nanowire exhibits a reduced parameter space where topology exists compared to the planar structure.

This is also illustrated by the effective parameters, shown in the right panels of Fig.~\ref{FigSM7}. We note that the exchange field exceeds the superconducting gap for various $V_{\rm tg}$ values. Some of these crossings are correlated to a dip in $\Delta_{\rm eff}$, indicating that the wavefunction is not proximitized by the superconductor and the system remains in the trivial regime. This is also illustrated by the color lines in the left panels of Fig.~\ref{FigSM7}.

In Fig.~\ref{FigSM8} we show the dispersion relation (left panels) and the density of states (right panels) for the same cases shown before. We have chosen parameters deep in a topological regime shown in Fig.~\ref{FigSM7}, when present. Notably, the superconducting gap of the wire is significantly reduced compared to the planar structure, see Fig.~\ref{Fig3} in the main text. Additionally, the gap appears to be soft, with many subgap states close to the Fermi level. These states are an obstacle towards applications and the demonstration of Majorana non-abelian properties. In general, softening of the gap can be attributed to two main effects: the presence of poorly proximitized subgap states in the semiconductor, and back-action of the SM-FI on the superconductor that suppress the pairing. Notice however that, while both effects can be identified in the nanowire case, the softening of the gap in this case can be mainly attributed to states localized in the parent superconductor (black lines in the left column). This suggests a stronger back-action of the FI and SM on the SC. This effect appears negligible in the 2DEG case. 

\begin{figure}
    \centering
    \includegraphics[width=1\columnwidth]{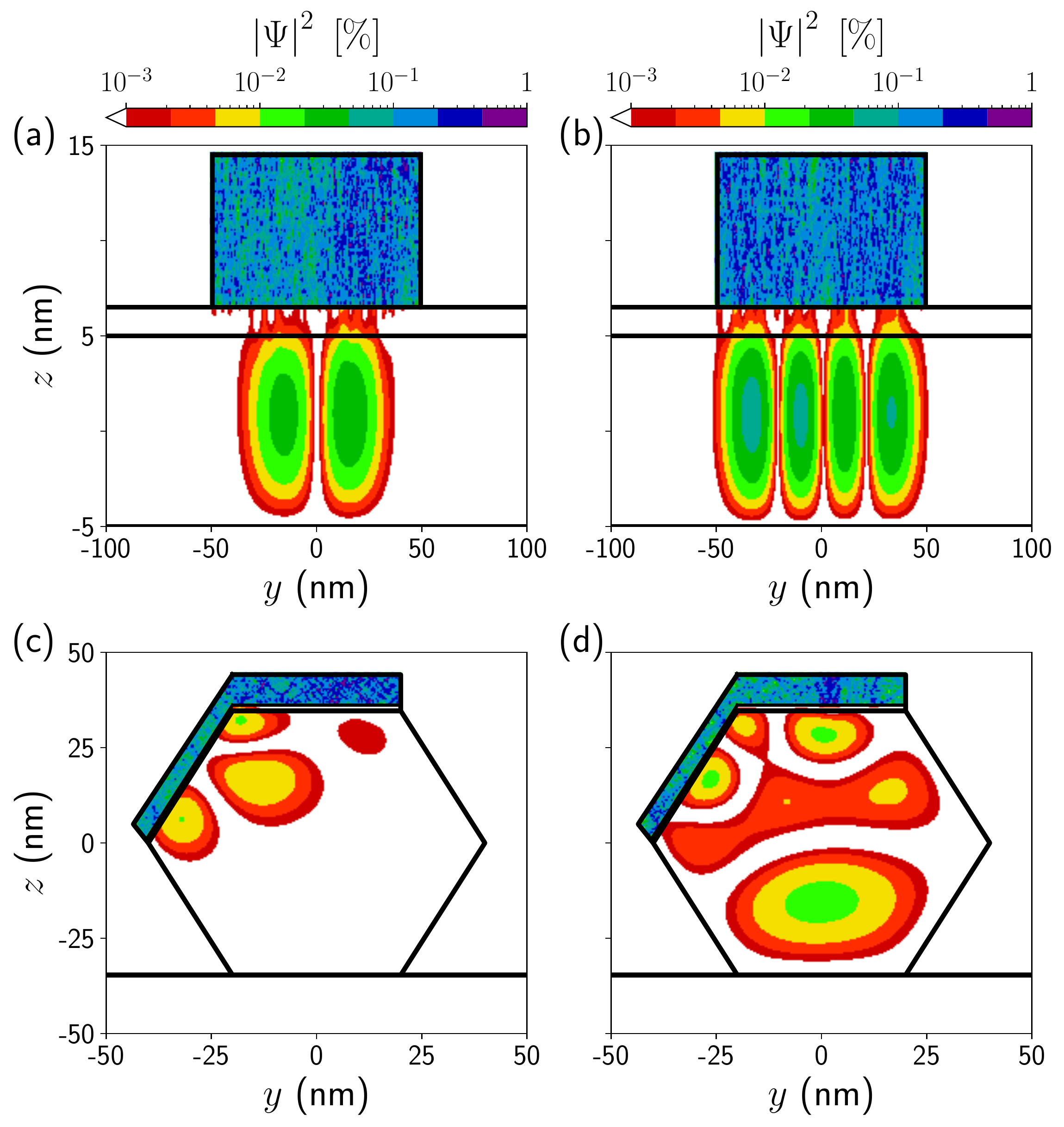}
    \caption{\textbf{Wavefunction profiles comparison between both geometries.} Wavefunction profile of the lowest-energy state in the 2DEG device (a) in a non-trivial topological phase close to pinch-off, $V_{\rm tg}=-850$ mV, and (b) in a different non-trivial phase after several subbands are populated in the wire, $V_{\rm tg}=-350$ mV. We take $d_{\rm FI}=1.5$ nm, and the rest of parameters are the same as in Fig. \ref{Fig1}. In (c,d), we show the same profile but for the wire device, also (c) in a non-trivial topological regime close to pinch-off, $V_{\rm bg}=-4$ V, and (d) in a different non-trivial phase but after several subbands are occupied in the wire, $V_{\rm bg}=-1$ V. For these two ones, the parameters are the same as in Fig.~\ref{FigSM1}.}
    \label{FigSM9}
\end{figure}

The main difference between the planar structure, presented in Fig.~\ref{Fig1} of the main text, and the hexagonal wire, Fig.~\ref{FigSM6} can be understood by looking at the wavefunction profiles. We show two examples of the wavefunction profile in Fig.~\ref{FigSM9} for the two geometries considered. The four cases correspond to the lowest-energy state in a topological regime. In the 2DEG geometry (a,b), the wavefunction is well localized below the SC stripe with a regular nodes distribution, top panels in Fig.~\ref{FigSM9}. This is a consequence of the strong vertical confinement imposed by the thin SM layer. In contrast, the wavefunction in the wire device, bottom panels in Fig.~\ref{FigSM9}, spreads across the whole cross section of the wire in some cases [Fig.~\ref{FigSM9}(d)], having a significant weight at positions several nm away from the SM-FI interface. The reduced localization at the interface and the irregular distribution affects the value of the effective superconducting pairing and exchange potential, resulting in the commented reduced topological regions, minigap, and the irregular distribution of the trivial and topological phases in parameter space.

\end{document}